\documentclass[pdflatex,sn-mathphys-num]{sn-jnl}

\usepackage{graphicx}%
\usepackage{multirow}%
\usepackage{amsmath,amssymb,amsfonts}%
\usepackage{amsthm}%
\usepackage{mathrsfs}%
\usepackage[title]{appendix}%

\usepackage{textcomp}%
\usepackage{manyfoot}%
\usepackage{booktabs}%
\usepackage{algorithm}%
\usepackage{algorithmicx}%
\usepackage{algpseudocode}%
\usepackage{listings}%
\usepackage{caption}
\usepackage{array}
\usepackage{pdflscape}
\usepackage{siunitx}

\usepackage[table]{xcolor}
\usepackage{float}

\usepackage{longtable}
\usepackage{geometry}
\usepackage{enumitem}
\usepackage{epstopdf}      
\usepackage{subcaption} 
\newcolumntype{C}{S[table-format=1.3, table-number-alignment=center]}

\theoremstyle{thmstyleone}%

\theoremstyle{thmstyletwo}%

\theoremstyle{thmstylethree}%

\begin{document}

\title[Article Title]{When Stein-Type Test Detects Equilibrium Distributions of Finite N-Body Systems}

\author[1,2,3]{\fnm{Jae Wan} \sur{Shim}}\email{jae-wan.shim@kist.re.kr}

\affil[1]{\orgdiv{Extreme Materials Research Center}, \orgname{Korea Institute of Science and Technology}, \orgaddress{\street{5 Hwarang-ro 14-gil, Seongbuk}, \city{Seoul}, \postcode{02792},  \country{Republic of Korea}}}

\affil[2]{\orgdiv{Climate and Environmental Research Institute}, \orgname{Korea Institute of Science and Technology}, \orgaddress{\street{5 Hwarang-ro 14-gil, Seongbuk}, \city{Seoul}, \postcode{02792},  \country{Republic of Korea}}}

\affil[3]{\orgdiv{Division of AI-Robotics}, \orgname{KIST Campus, University of Science and Technology}, \orgaddress{\street{5 Hwarang-ro 14-gil, Seongbuk}, \city{Seoul}, \postcode{02792},  \country{Republic of Korea}}}

\abstract{
Starting from the probability distribution of finite N-body systems, which maximises the Havrda--Charv\'at entropy, we build a Stein-type goodness-of-fit test. The Maxwell--Boltzmann distribution is exact only in the thermodynamic limit, where the system is composed of infinitely many particles as N approaches infinity. For an isolated system with a finite number of particles, the equilibrium velocity distribution is compact and markedly non-Gaussian, being restricted by the fixed total energy. Using Stein's method, we first obtain a differential operator that characterises the target density.  Its eigenfunctions are symmetric Jacobi polynomials, whose orthogonality yields a simple, parameter-free statistic. Under the null hypothesis that the data follows the finite-N distribution, the statistic converges to a chi-squared law, so critical values are available in closed form. Large-scale Monte Carlo experiments confirm exact size control and give a clear picture of the power. These findings quantify how quickly a finite system approaches the classical limit and provide a practical tool for testing kinetic models in regimes where normality cannot be assumed.
}


\maketitle

\section{Introduction}\label{sec:intro}
The Maxwell--Boltzmann distribution is a foundational result of statistical physics. Maxwell first obtained the familiar Gaussian velocity distribution by invoking symmetry arguments, whereas Boltzmann later solidified its physical relevance dynamically: his $H$-theorem shows that an entropy functional is non-decreasing in time along the Boltzmann evolution and that the Maxwellian corresponds to equilibrium. Although this celebrated result predicts an apparent monotonic growth of entropy, Boltzmann himself emphasised that the approach to equilibrium is statistical rather than absolute, occurring with overwhelming probability rather than certainty \citep{boltzmann1898ueber, shim2024commented}.

Jaynes later reached the same distribution from a static, inference-based standpoint. His Principle of Maximum Entropy asserts that, given incomplete information, the least biased probability law is the one that maximises (Shannon) entropy subject to the known constraints \citep{jaynes1957information}. In the thermodynamic limit ($N\to\infty$), maximising Shannon entropy under a constraint on the average energy naturally reproduces the classical Maxwell--Boltzmann distribution.

For an isolated system with a finite number of particles, however, the exact conservation of total energy restricts the accessible phase space. The equilibrium single-particle velocity distribution consequently acquires compact support and a flatter central peak relative to the canonical Gaussian, approaching the Gaussian only as $N\to\infty$. While this microcanonical marginal can be derived directly from mechanics, reproducing it via a maximum entropy principle with simple constraints motivates an entropic functional beyond Shannon's form. Although \citet{shore1980axiomatic} argued that the Kullback--Leibler relative entropy is uniquely fixed by a set of consistency axioms, Uffink~\cite{uffink1995can} later showed that the Shore--Johnson axioms are overly restrictive; a one-parameter family of R\'enyi entropies can also be made to satisfy suitably weakened versions of these requirements, demonstrating that Shannon's measure is not unique.

The present work explores the statistics of such finite-$N$ equilibrium velocity distributions. The parameter $N$ functions not as a mere particle count but as an effective statistical index quantifying deviations from Gaussian behaviour. Systems exhibiting long-range correlations, non-ergodicity, or other non-extensive features can often be mapped onto an effective finite-$N$ description. Determining whether empirical data follow a finite-$N$ distribution can therefore provide a useful diagnostic of the underlying non-extensive character of the system.

\subsection{Finite-$N$ Distribution and Exact Geometry}
We adopt the Havrda--Charv\'at entropy (also known as Tsallis entropy) as the information measure for finite, isolated microcanonical systems. Maximising this entropy subject to normalisation and the fixed total kinetic energy yields the compactly supported single-particle velocity distribution \citep{Havrda1967, tsallis1988possible, shim2020entropy, shim2023generalized}:
\begin{equation}
f_{D,N}(\mathbf v)=
\frac{\Gamma\!\left(\frac{DN}{2}\right)}
     {\Gamma\!\left(\frac{D(N-1)}{2}\right)\,(\pi U)^{D/2}}
     \left(1-\frac{\|\mathbf v\|^{2}}{U}\right)^{\frac{D(N-1)-2}{2}}_{+},
\qquad
N>1+\frac{2}{D},
\label{eq:HC_general}
\end{equation}
where \(\Gamma(\cdot)\) denotes the Gamma function, \(\mathbf v\in\mathbb R^{D}\) is the velocity of a tagged particle, and \((x)_{+}=\max(0,x)\). Writing the fixed total kinetic energy of the $N$-particle system as
\(E=\frac{m}{2}\sum_{i=1}^{N}\|\mathbf v_i\|^{2}\),
we set \(U:=2E/m\), so that \(mU/2=E\). Physically, \(U\) is related to the absolute temperature \(T\) via equipartition as \(\frac{mU}{2}= \frac{DN}{2}kT\), where \(k\) is the Boltzmann constant.

\paragraph{Location--scale extension.}
For any $\mu\in\mathbb R^D$ and $\sigma>0$, define the reparametrised density by the change of variables
$\mathbf z=(\mathbf v-\boldsymbol\mu)/\sigma$, i.e.
\[
f_{D,N}^{(\boldsymbol\mu,\sigma)}(\mathbf v)
:=\frac{1}{\sigma^{D}}\,f_{D,N}\!\left(\frac{\mathbf v-\boldsymbol\mu}{\sigma}\right).
\]
The Jacobian factor $\sigma^{-D}$ ensures normalisation $\int f_{D,N}^{(\boldsymbol\mu,\sigma)}\,d\mathbf v=1$.
Moreover, by symmetry of $f_{D,N}$ we have $\mathbb E[\mathbf Z]=\mathbf 0$, hence
$\mathbb E[\mathbf V]=\boldsymbol\mu$ and
\[
\mathrm{Cov}(\mathbf V)=\sigma^{2}\,\mathrm{Cov}(\mathbf Z),
\qquad\text{in particular}\quad
\mathbb E\!\left[\|\mathbf V-\boldsymbol\mu\|^{2}\right]
=\sigma^{2}\,\mathbb E\!\left[\|\mathbf Z\|^{2}\right].
\]
In the one-dimensional normalisation used below (Eq.~\eqref{eq:p_N}), $\mathbb E[Z^2]=1$, so
$\mathbb E[V]=\mu$ and $\mathbb E[(V-\mu)^2]=\sigma^2$.

The choice of this distribution is not arbitrary. As we will rigorously demonstrate, the probability functional derived from the exact geometry of the finite phase space (hypersphere surface ratios) is mathematically identical to the one derived from the Havrda--Charv\'at entropy structure without Stirling's approximation. This establishes a precise link between the microcanonical geometry of finite systems and generalised information theory.

\subsection{Theoretical Justification: Consistency and Entropy Equivalence}

A derivation of the finite-\(N\) distribution in Eq.~\eqref{eq:HC_general} relies on the maximisation of the Havrda--Charv\'at entropy. A common objection to using non-Shannon entropies is the claim that only the Shannon entropy satisfies the consistency axioms for statistical inference, particularly system independence \citep{shore1980axiomatic}. However, \citet{uffink1995can} demonstrated that this uniqueness claim relies on overly restrictive assumptions. He showed that, once these assumptions are suitably weakened, a broader class of generalised entropies, including the R\'enyi entropies, can satisfy the relevant consistency requirements for inductive inference. In the microcanonical setting considered here, the finite-$N$ single-particle distribution is also obtained directly from the standard microcanonical postulate: the $N$-particle velocity vector is uniformly distributed on the fixed-energy hypersphere, and integrating out $N-1$ particles yields Eq.~\eqref{eq:HC_general}. The Havrda--Charv\'at maximisation therefore provides an information-theoretic route to the same exact microcanonical marginal, rather than an ad hoc modelling choice.

Here, we demonstrate that the Havrda--Charv\'at entropy \(S_\alpha\) and the R\'enyi entropy \(H_\alpha\) \citep{renyi1961measures} are monotonically related. Consequently, maximising \(S_\alpha\) is mathematically equivalent to maximising \(H_\alpha\), and thus inherits the statistical consistency justified by \citet{uffink1995can} under the same constraints.

Let \(P = \{p_1, \dots, p_n\}\) be a probability distribution. For a parameter \(\alpha \in \mathbb{R} \setminus \{1\}\), the two entropies are defined as:
\begin{align}
    H_\alpha(P) &= \frac{1}{1-\alpha} \ln \left( \sum_{i=1}^n p_i^\alpha \right), \label{eq:Renyi} \\
    S_\alpha(P) &= \frac{1}{1-\alpha} \left( \sum_{i=1}^n p_i^\alpha - 1 \right). \label{eq:Tsallis}
\end{align}
From these two equations, we obtain
\begin{equation}
    H_\alpha(P) = \frac{1}{1-\alpha} \ln \left[ 1 + (1-\alpha)S_\alpha(P) \right].
\end{equation}
To show that \(H_\alpha\) is a strictly increasing function of \(S_\alpha\), we differentiate \(H_\alpha\) with respect to \(S_\alpha\):
\[
\begin{aligned}
    \frac{d H_\alpha}{d S_\alpha} 
    &= \frac{1}{1 + (1-\alpha)S_\alpha(P)}
     = \frac{1}{\sum_{i=1}^n p_i^\alpha}.
\end{aligned}
\]
For $\alpha>0$ we have $\sum_i p_i^\alpha>0$ whenever $p_i\ge 0$ and $\sum_i p_i=1$. For $\alpha<0$ the same conclusion holds provided $p_i>0$ for all $i$. Therefore,
\begin{equation}
    \frac{d H_\alpha}{d S_\alpha} = \frac{1}{\sum_{i=1}^n p_i^\alpha} > 0,
\end{equation}
which proves that \(H_\alpha\) is a strictly monotonically increasing function of \(S_\alpha\).
 
Since the objective function is merely transformed by a monotonic mapping, the probability distribution \(P^*\) that maximises the Havrda--Charv\'at entropy \(S_\alpha\) under a given set of constraints is identical to the distribution that maximises the R\'enyi entropy \(H_\alpha\) under the same constraints. Thus, the finite-\(N\) distribution \(p_N(x)\) derived in this work rests on a statistically consistent foundation as discussed by \citet{uffink1995can}.

\subsection{Specialising to one dimension}
Throughout this paper, we analyse a \emph{single} velocity component, so we set \(D=1\). We further adopt the conventional microcanonical normalisation
\[
\sum_{i=1}^{N} v_i^{2}=N,
\]
which merely fixes the velocity unit (and thereby sets the associated temperature scale via equipartition). With this choice, the total kinetic energy is
\[
\frac{mU}{2}=\frac{m}{2}\sum_{i=1}^{N} v_i^{2}
           =\frac{mN}{2},
\]
so we identify the energy scale with the particle count,
\[
D=1,\qquad U=N,
\]
with the inherited condition \(N>3\) from Eq.~\eqref{eq:HC_general}. Substituting \(D=1\) and \(U=N\) into Eq.~\eqref{eq:HC_general} yields the one-dimensional finite-\(N\) distribution
\begin{equation}
p_{N}(x)=
\frac{\Gamma\!\bigl(\frac{N}{2}\bigr)}
     {\sqrt{\pi N}\,\Gamma\!\bigl(\frac{N-1}{2}\bigr)}
     \Bigl(1-\tfrac{x^{2}}{N}\Bigr)^{\!\frac{N-3}{2}}_{+},
\qquad |x|\le\sqrt{N}.
\label{eq:p_N}
\end{equation}
This density has finite support and a \emph{flatter} central peak than the Gaussian, converging to it only as \(N\to\infty\).

This derivation makes explicit how the parameters in the Havrda--Charv\'at framework map onto the physical one-dimensional velocity component and clarifies the origin of the exponent \((N-3)/2\) and the hard cutoff at \(\sqrt{N}\).


\section{Derivation of the Likelihood Ratio for a Typical Configuration}
\label{sec:derivation_typical}

We quantify how reliably a Gaussian sample can be \emph{distinguished from} the finite-\(N\) law
by evaluating the likelihood ratio on the \emph{typical set of} \(p_N\).
Equivalently, we estimate the probability that a draw from the Gaussian benchmark is mapped into
a typical configuration of the finite microcanonical surface.

\subsection{Exact Likelihoods}

Let \(\mathbf{x} = (x_1,\dots,x_n)\) denote an i.i.d.\ sample of size \(n\) drawn from the one-dimensional marginal \(p_N(x)\) under \(H_0\) and from \(p_\infty(x)\) under \(H_1\).

\paragraph{\textbf{Null hypothesis} \(H_0\) (finite \(N\)).}
\[
  L_N(\mathbf{x})
    \;=\;
    C_N^{\,n}\,
    \exp\!\Biggl[
        \frac{N-3}{2}
        \sum_{i=1}^{n}
        \ln\!\Bigl(1-\frac{x_i^{2}}{N}\Bigr)
    \Biggr],
    \qquad
    C_N := \frac{\Gamma\!\bigl(\tfrac{N}{2}\bigr)}
                 {\sqrt{N\pi}\,\Gamma\!\bigl(\tfrac{N-1}{2}\bigr)} .
\]

\paragraph{\textbf{Alternative hypothesis} \(H_1\) (Gaussian).}
\[
  L_\infty(\mathbf{x}) \;=\; (2\pi e)^{-\,\frac{n}{2}} \quad \text{under the condition of } \sum x_i^2 = n.
\]
Here we evaluate the Gaussian likelihood on the typical shell where \(n^{-1}\sum_{i=1}^n x_i^2 \approx 1\), so we replace \(\sum x_i^2\) by \(n\) at leading order.

\paragraph{\textbf{Likelihood ratio (in favour of the Gaussian)}.}
\[
  \Lambda_N(\mathbf{x}) \;:=\; \frac{L_\infty(\mathbf{x})}{L_N(\mathbf{x})}.
\]

\subsection{Typical-State Approximation}

For large \(n\) the empirical average
\(
  n^{-1}\sum_{i=1}^{n}\ln\bigl(1-x_i^{2}/N\bigr)
\)
converges almost surely to its expectation with respect to \(p_N\) by the law of large numbers:
\[
  \mathbb{E}_{p_N}\!\left[
     \ln\!\bigl(1-x^{2}/N\bigr)
  \right]
  \;=\;
  \Delta\psi_N,
  \qquad
  \Delta\psi_N :=
    \psi\!\bigl(\tfrac{N-1}{2}\bigr)
    - \psi\!\bigl(\tfrac{N}{2}\bigr) ,
\]
with \(\psi\) the digamma function.  Inside the typical set we therefore
replace the fluctuating sum by its deterministic limit:
\[
  \sum_{i=1}^{n}\ln\!\Bigl(1-\tfrac{x_i^{2}}{N}\Bigr)
  \;\longrightarrow\;
  n\,\Delta\psi_N .
\]

\subsection{Closed-Form Result}

Inserting the above into \(L_N\) into the likelihood ratio, we obtain the \textit{typical} likelihood ratio
\[
  \Lambda_N^{\mathrm{typ}}
    \;=\;
    \frac{(2\pi e)^{-\,\frac{n}{2}}}
         {C_N^{\,n}\,
          \exp\!\bigl[\,\tfrac{n}{2}(N-3)\,\Delta\psi_N\bigr]} .
\]
Employing the explicit Gamma-function form of \(C_N\) we finally arrive at
\begin{equation}
\boxed{%
  \Lambda_N^{\mathrm{typ}}
  \;=\;
  \Biggl[
    \sqrt{\frac{N}{2e}}\,
    \frac{\Gamma\!\bigl(\tfrac{N-1}{2}\bigr)}
         {\Gamma\!\bigl(\tfrac{N}{2}\bigr)}
    \exp\!\Bigl\{
       -\frac{N-3}{2}\,
        \bigl[
          \psi\!\bigl(\tfrac{N-1}{2}\bigr)
          -\psi\!\bigl(\tfrac{N}{2}\bigr)
        \bigr]
    \Bigr\}
  \Biggr]^{\!n}}
\label{eq:typ_likelihood_ratio_rewrite}
\end{equation}

\subsection{Interpretation as a Probability}

Denote by \(A_N\) the typical set of \(p_N\), chosen so that \(\int_{A_N}p_N(\mathbf{x})\,d\mathbf{x}\approx1\).
Here \(p_N(\mathbf{x})=\prod_{i=1}^n p_N(x_i)\) and \(p_\infty(\mathbf{x})=\prod_{i=1}^n p_\infty(x_i)\).
Consider the typical-set decision rule that \emph{accepts} \(H_0\) whenever \(\mathbf{x}\in A_N\).
Then the probability that a Gaussian draw falls into \(A_N\) (i.e.\ the Type~II error of this rule) is
\[
  \mathcal{P}
    \;=\;
    \int_{A_N} p_\infty(\mathbf{x})\,d\mathbf{x}
    \;=\;
    \int_{A_N} p_N(\mathbf{x})\,\Lambda_N(\mathbf{x})\,d\mathbf{x}.
\]
Because \(\Lambda_N(\mathbf{x})\) is nearly constant over \(A_N\) and coincides with \(\Lambda_N^{\mathrm{typ}}\), the integral reduces to
\begin{equation}
  \mathcal{P} \;\approx\; \Lambda_N^{\mathrm{typ}} .
\label{eq:prob_typical}
\end{equation}

\subsection{Consistency with Sanov's Theorem}

The exact Kullback--Leibler divergence between \(p_N\) and \(p_\infty\) (see Appendix~\ref{sec:derivation_DKL}) is
\[
  D_{\mathrm{KL}}(p_N \,\|\, p_\infty)
    \;=\;
      \ln C_N
      + \tfrac{1}{2}\bigl(1+\ln 2\pi\bigr)
      + \tfrac{N-3}{2}\,\Delta\psi_N .
\]
Combining this with Eq.~\eqref{eq:typ_likelihood_ratio_rewrite} gives
\[
  \Lambda_N^{\mathrm{typ}}
    \;=\;
    \exp\!\bigl[-\,n\,D_{\mathrm{KL}}(p_N \,\|\, p_\infty)\bigr].
\]
This matches Sanov's large-deviation form~\citep{sanov1961probability, cover1991elements} for the probability that an i.i.d.\ sample from \(p_\infty\)
has an empirical law close to \(p_N\), namely \(\mathcal{P}\asymp \exp(-n D_{\mathrm{KL}}(p_N\|p_\infty))\).
Hence, the finite-\(N\) geometry governed by the Havrda--Charv\'at entropy naturally reproduces the Sanov rate
without recourse to Stirling or thermodynamic approximations.


\section{Stein Characterisation for Finite-\texorpdfstring{$N$}{N} Laws}
\label{sec:stein_method}

Having established in previous sections how the finite-\(N\) density \(p_N(x)\) arises and how it departs from the Gaussian benchmark, we now develop a \emph{Stein-type goodness-of-fit statistic} that exploits those departures in a systematic way. Our work is built within the Stein paradigm first outlined by Stein~\cite{stein1972} and later elaborated by Chen \emph{et al.}~\cite{chen2010normal}, but it is carefully adjusted to the finite support and algebraic boundary decay, rather than exponential tails, of \(p_N\).

\subsection{General Stein Framework}

Let \(P\) be a probability law supported on the interval \([a,b]\subset\mathbb{R}\).
A linear map \(\mathcal{A}\) is said to be a \emph{Stein operator} for \(P\) if
\[
   \mathbb{E}_{P}\!\bigl[(\mathcal{A}f)(X)\bigr] = 0
   \quad\text{for every } f\in\mathcal{F},
\]
where \(\mathcal{F}\) is a suitable class of test functions that are absolutely continuous on \([a,b]\) with an
integrable derivative. For one-dimensional densities \(p(x)\) that vanish at the boundaries, Goldstein \& Reinert \cite{goldstein2013stein} showed that a first-order operator of the form
\begin{equation}
  (\mathcal{A}f)(x)
    \;=\;
    s(x)f'(x)
    +\bigl[s'(x)+s(x)(\ln p)'\bigr]f(x)
\label{eq:stein_first_order_general}
\end{equation}
is sufficient, provided the weight \(s(x)\) satisfies \(s(a)=s(b)=0\).

\subsection{Stein Operator for the Finite-$N$ Density}

For the compact-support law \(p_N(x)\) in Eq.~\eqref{eq:p_N}, the natural boundary points are \(a=-\sqrt{N}\) and \(b=\sqrt{N}\). For \(N>3\) we have \(p_N(\pm\sqrt{N})=0\), so the boundary terms vanish. Choosing the vanishing weight
\begin{equation}
  s_N(x)\;:=\;1-\frac{x^{2}}{N},
\end{equation}
and writing \(\alpha:=(N-3)/2\) for brevity, we obtain
\[
  (\ln p_N)'(x)
    \;=\;
    \frac{d}{dx}\left[\alpha\ln\!\left(1-\frac{x^{2}}{N}\right)\right]
    \;=\;
    -\frac{2\alpha x}{N\left(1-\frac{x^{2}}{N}\right)}.
\]
Substituting \(s_N\) and \((\ln p_N)'\) into Eq.~\eqref{eq:stein_first_order_general} yields
\[
  s_N'(x)=-\frac{2x}{N},
  \qquad
  s_N(x)(\ln p_N)'(x)=-\frac{2\alpha x}{N},
\]
and hence the coefficient of \(f(x)\) is
\[
  b_N(x)
    \;:=\;
    s_N'(x)+s_N(x)(\ln p_N)'(x)
    =-\frac{N-1}{N}\,x.
\]
Hence the finite-\(N\) Stein operator is
\begin{equation}
  \boxed{%
    (\mathcal{A}_N f)(x)
    \;=\;
    \left(1-\frac{x^{2}}{N}\right)f'(x)
    \;-\;\frac{N-1}{N}\,x\,f(x)},
  \qquad |x|\le\sqrt{N}.
\label{eq:stein_operator_finiteN}
\end{equation}
As \(N\to\infty\), \(s_N(x)\to1\) and the coefficient of \(f(x)\) tends to \(-x\), so \(\mathcal{A}_N\) smoothly approaches the Ornstein--Uhlenbeck operator \(f'(x)-xf(x)\) that characterises the standard normal law, in agreement with the thermodynamic limit discussed earlier.

\subsection{Rescaling to the Jacobi Form}

To exploit the orthogonality of classical polynomials it is convenient to map the support \([-\,\sqrt{N},\sqrt{N}\,]\) onto the reference interval \([-1,1]\).  Setting \(y:=x/\sqrt{N}\) and \(g(y):=f(\sqrt{N}y)\) leads to the rescaled operator
\[
  (\widetilde{\mathcal{A}}_N g)(y)
  :=\sqrt{N}\,(\mathcal{A}_N f)(\sqrt{N}y)
  =(1-y^{2})g'(y)\;-\;(N-1)\,y\,g(y),
  \qquad y\in[-1,1].
\]

Let \(w_N(y)\propto(1-y^2)^{(N-3)/2}\) denote the target weight on \([-1,1]\).
Although \(\widetilde{\mathcal A}_N\) is first order, setting \(g=h'\) induces the Jacobi Sturm--Liouville operator
\[
(\mathcal L_N h)(y):=(\widetilde{\mathcal A}_N h')(y)=(1-y^2)h''(y)-(N-1)y h'(y)=\frac{1}{w_N(y)}\frac{d}{dy}\Bigl((1-y^2)\,w_N(y)\,h'(y)\Bigr).
\]
Since \(\mathcal L_N\) is self-adjoint in \(L^2([-1,1],w_N)\), its orthogonal polynomial eigenfunctions are the symmetric Jacobi polynomials.

The first few modes correspond to variance, kurtosis, and higher-order corrections made explicit in Section~\ref{sec:test}, thereby providing an explicit orthogonal basis for finite-\(N\) goodness-of-fit testing.

\subsection{Implications for Goodness-of-Fit Testing}

Given an i.i.d.\ sample \(\mathbf{x}\), one may estimate the Stein discrepancy
\(
  \sup_{f\in\mathcal{F}}\bigl|\tfrac{1}{n}\sum_{i=1}^{n}(\mathcal{A}_N
  f)(x_i)\bigr|
\)
by truncating the Jacobi-polynomial expansion at a modest order, typically \(m=4\) to \(6\).  As we show in the numerical experiments, such a truncation already captures the leading finite-size deviations and delivers high detection power for \(N\lesssim10\) with a few hundred observations, while larger \(N\) require longer samples as finite-size effects recede.

In short, the operator in~\eqref{eq:stein_operator_finiteN} supplies a Stein toolkit for finite-\(N\) systems. It admits closed-form analysis and converges smoothly to the classical Gaussian operator as \(N\to\infty\).

\section{Stein-Type Test Statistic}\label{sec:test}

The symmetric Jacobi polynomials form an orthogonal basis under the symmetric Beta-type (Jacobi) weight on \([-1,1]\), which is convenient for constructing Stein-type goodness-of-fit tests\citep{schoutens2001orthogonal}. In this section, we derive the test statistic.
Throughout this section we set \(\alpha=(N-3)/2\), so that \(N-1=2(\alpha+1)\).

\subsection{Jacobi polynomials and the Beta weight}
For \(\alpha > -1\), the symmetric Jacobi polynomials \(P_k^{(\alpha,\alpha)}(y)\) (\(k=0,1,\dots\)) form an orthogonal basis on \([-1,1]\) with respect to the normalised symmetric Beta-type (Jacobi) weight:
\begin{equation}
w_\alpha(y) = \frac{1}{B(\frac{1}{2}, \alpha+1)} (1-y^2)^\alpha 
            = \frac{\Gamma(\alpha+\frac{3}{2})}{\sqrt{\pi}\,\Gamma(\alpha+1)} (1-y^2)^\alpha, \quad |y|\le 1.
\end{equation}
The orthogonality condition is given by
\[
\int_{-1}^{1} P_k^{(\alpha,\alpha)}(y)\, P_\ell^{(\alpha,\alpha)}(y)\, w_\alpha(y)\,dy = 0 \quad (k\ne\ell).
\]
Using the Gamma-function duplication formula \(\Gamma(2z) = \pi^{-1/2} 2^{2z-1}\Gamma(z)\Gamma(z+1/2)\) with \(z=\alpha+1\), the normalising constant can be rewritten in a form that highlights the connection to the symmetrised Beta law:
\[
w_\alpha(y) = \frac{\Gamma(2\alpha+2)}{2^{2\alpha+1}\Gamma(\alpha+1)^{2}} (1-y^2)^\alpha.
\]

\subsection{Differential equations for symmetric Jacobi polynomials}
The symmetric Jacobi polynomials are eigenfunctions of a Sturm--Liouville operator. Specifically, \(P_k^{(\alpha,\alpha)}(y)\) satisfies the differential equation:
\begin{equation}
(1-y^{2})P_k''(y) - 2(\alpha+1)y P_k'(y) + k(k+2\alpha+1)P_k(y) = 0.
\label{eq:JacobiODE}
\end{equation}
This second-order ODE is crucial because its differential part, \((1-y^2)\frac{d}{dy} - 2(\alpha+1)y\), matches the structure of our unit-support Stein operator \(\widetilde{\mathcal A}_N\) derived in the previous section (with \(N-1 = 2\alpha+2\)).

\subsection{Action of \(\widetilde{\mathcal A}_N\) on the test basis}
To construct the test statistic, we choose a basis of test functions derived from the Jacobi polynomials. Let the test functions be the shifted Jacobi polynomials:
\[
g_k(y) := P_{k-1}^{(\alpha+1,\alpha+1)}(y), \qquad k=1,2,\ldots
\]
We utilise the classical derivative identity for Jacobi polynomials \citep{szego1939}:
\begin{equation}
\frac{d}{dy} P_k^{(\alpha,\alpha)}(y) = \frac{k+2\alpha+1}{2} P_{k-1}^{(\alpha+1,\alpha+1)}(y).
\label{eq:dJacobi}
\end{equation}
From \eqref{eq:dJacobi}, we can express the test function \(g_k\) as a scaled derivative of the target orthogonal polynomial:
\begin{equation}
g_k(y) = \frac{2}{k+2\alpha+1} \frac{d}{dy} P_k^{(\alpha,\alpha)}(y).
\label{eq:g_k_relation}
\end{equation}
Applying the unit-support Stein operator \(\widetilde{\mathcal A}_N g(y) = (1-y^2)g'(y) - 2(\alpha+1)y g(y)\) to \(g_k\) and utilising the ODE \eqref{eq:JacobiODE} for \(P_k^{(\alpha,\alpha)}\), we obtain a remarkable algebraic simplification:
\[
\begin{aligned}
(\widetilde{\mathcal A}_N g_k)(y)
&= (1-y^2)g_k'(y) - 2(\alpha+1)y g_k(y) \\
&= \frac{2}{k+2\alpha+1} \left[ (1-y^2)\frac{d^2}{dy^2}P_k^{(\alpha,\alpha)}(y) - 2(\alpha+1)y \frac{d}{dy}P_k^{(\alpha,\alpha)}(y) \right] \\
&= \frac{2}{k+2\alpha+1} \left[ -k(k+2\alpha+1) P_k^{(\alpha,\alpha)}(y) \right] \\
&= -2k\, P_k^{(\alpha,\alpha)}(y).
\end{aligned}
\]
Thus, the Stein operator maps the shifted basis \(g_k\) directly onto the orthogonal basis of the target density:
\begin{equation}
\boxed{ (\widetilde{\mathcal A}_N g_k)(y) = -2k\, P_k^{(\alpha,\alpha)}(y), \quad k \ge 1. }
\label{eq:Saction-correct}
\end{equation}

\subsection{Variance and Normalisation of the Stein Basis}
Since the image of the Stein operator is proportional to the orthogonal polynomials \(P_k^{(\alpha,\alpha)}\), the resulting components are mutually uncorrelated under the null hypothesis. The variance of the \(k\)-th component is:
\[
\sigma_k^2 := \mathbb{E}_{w_\alpha} \bigl[ (\widetilde{\mathcal A}_N g_k)(Y)^2 \bigr] = 4k^2 \int_{-1}^{1} \bigl[ P_k^{(\alpha,\alpha)}(y) \bigr]^2 w_\alpha(y)\, dy.
\]
Using the standard norm of Jacobi polynomials and the normalisation constant of \(w_\alpha\), we obtain the closed-form expression:
\begin{equation}
\sigma_k^2 = 4k^2 \cdot \frac{\Gamma(\alpha+\frac{3}{2})}{\sqrt{\pi}\Gamma(\alpha+1)} \cdot \frac{2^{2\alpha+1}}{2k+2\alpha+1} \frac{\Gamma(k+\alpha+1)^2}{k!\,\Gamma(k+2\alpha+1)}.
\label{eq:sigma-closed-form}
\end{equation}
The values are provided in Table~\ref{tab:sigma-values}.
We define the orthonormalised Stein basis functions as:
\[
\psi_k(y) := \frac{(\widetilde{\mathcal A}_N g_k)(y)}{\sigma_k} = -\frac{2k}{\sigma_k} P_k^{(\alpha,\alpha)}(y).
\]

\begin{table}[t]
\centering
\caption{Normalisation constants $\sigma_k$ for the orthonormalised Stein basis functions with particle number \textbf{$N=5$} ($\alpha=1$). The analytical values are computed via Eq.~\eqref{eq:sigma-closed-form}.}
\begin{tabular}{c c | c c}
\toprule
Order $k$ & $\sigma_k$ & Order $k$ & $\sigma_k$ \\
\midrule
1 & 1.7889 & 6 & 7.0993 \\
2 & 3.2071 & 7 & 7.8416 \\
3 & 4.3818 & 8 & 8.5298 \\
4 & 5.3936 & 9 & 9.1736 \\
5 & 6.2897 & 10 & 9.7802 \\
\bottomrule
\end{tabular}
\label{tab:sigma-values}
\end{table}

\subsection{The Test Statistic}
The orthonormality condition \(\mathbb{E}[\psi_k(Y)\psi_\ell(Y)] = \delta_{k\ell}\) implies that for an i.i.d.\ sample \(Y_1, \dots, Y_n\) drawn from \(p_N\), the empirical coefficients
\begin{equation}
    \widehat{\mu}_k = \frac{1}{\sqrt{n}} \sum_{i=1}^n \psi_k(Y_i)
\end{equation}
are asymptotically independent standard normal variables.

In practice, we treat location and scale as nuisance parameters and therefore standardise the sample (enforcing zero mean and unit second moment) before applying the test. This standardisation corresponds to the location--scale extension and the induced Stein operator for the standardised variable. Furthermore, since both the target distribution $p_N$ and the Gaussian alternative are symmetric \emph{after this location--scale alignment}, the first three components ($k=1,2,3$) corresponding to mean, variance, and skewness provide no discrimination power. We therefore define the targeted test statistic over a selected set of modes \(\mathcal{K}\) (e.g., \(\mathcal{K}=\{4, \dots, m\}\)):
\begin{equation}
    T_{n, \mathcal{K}} = \sum_{k \in \mathcal{K}} \widehat{\mu}_k^2.
\end{equation}
Under the null hypothesis, \(T_{n, \mathcal{K}}\) converges in distribution to a chi-squared random variable \(\chi^2_d\), where the degrees of freedom \(d = |\mathcal{K}|\) denotes the cardinality of the mode set.


\section{Numerical Implementation and Simulation Results}
\label{sec:numerics}

This section details the implementation of the test, the layout of the Monte Carlo experiment, and
the comparison of the resulting power curves with the exponential rate predicted by Sanov's theorem.

\subsection{Implementation Details}

In implementation we work with the rescaled variable \(Y=X/\sqrt{N}\in[-1,1]\), which is a fixed change of units and does not affect the test; we report results in terms of \((N,n,m)\) throughout.
The orthonormal functions
\(\psi_k(y)\propto P_k^{(\alpha,\alpha)}(y)\) are built for
\(\alpha=(N-3)/2\).

\begin{itemize}
  \item \textbf{Recurrence.}  Symmetric Jacobi polynomials are generated by the three-term rule
        \[
          (k+1)(k+2\alpha+1)P_{k+1}^{(\alpha,\alpha)}(y)
          =(2k+2\alpha+1)(k+\alpha+1)\,y\,P_k^{(\alpha,\alpha)}(y)
          -(k+\alpha)(k+\alpha+1)\,P_{k-1}^{(\alpha,\alpha)}(y),
        \]
        with \(P_{0}=1\) and \(P_{1}=(\alpha+1)y\).
  \item \textbf{Normalisation.}  The constants \(\sigma_k\) are pre-computed from the closed form in Eq.~\eqref{eq:sigma-closed-form}.
\end{itemize}

\subsection{Simulation Setup}
We report results under two calibrations: an asymptotic \(\chi^2_d\) threshold and a Monte Carlo calibrated threshold obtained under \(H_0\) for each \((N,n,m)\).

\begin{itemize}
  \item \textbf{Null data (\(H_0\)).} Draw \(X\sim p_N(x)\) and evaluate the test on the rescaled variable \(Y=X/\sqrt{N}\in[-1,1]\).
  \item \textbf{Alternative data (\(H_1\)).} Draw \(X\sim \mathcal{N}(0,1)\) and evaluate the test on \(Y=X/\sqrt{N}\) (equivalently, \(Y\sim\mathcal{N}(0,1/N)\)).
\item \textbf{Grid.}
      \(\;N\in\{5,6,7,\ldots,20\},\;
        n\in\{10,20,30,\ldots,200\}\cup\{250,300,350,400,450,500\}.\)
  \item \textbf{Truncation.}
        We evaluated \(m \in \{4, 6, 8, 10\}\).
        Since both \(H_0\) and \(H_1\) are symmetric and mean- and variance-matched in the rescaled variable \(Y\), odd modes and low-order modes are redundant for discrimination.
        Accordingly, we used the even-mode set \(\mathcal{K}=\{4,6,\dots,m\}\).
        Note that additional modes may be retained depending on the specific application.
  \item \textbf{Repetitions.} Monte Carlo calibration used \(50{,}000\) replications under \(H_0\), and size/power estimation used \(20{,}000\) replications per setting, at the \(5\%\) level.
\end{itemize}

\subsection{Type I Error Control}

Table~\ref{tab:type1_error} reports empirical Type~I error under \(H_0\) using
calibrated (Calibr.) and theoretical \(\chi^2\) (Theoret.) critical values.
With calibration, the rejection rates remain close to the nominal level \(0.05\)
across the reported regimes, with the recommended truncation level \(m=4\) showing the most stable behavior.
In contrast, the theoretical cutoff exhibits visible finite-sample distortion:
it is conservative for \(N=20\) and \(n=10\) (e.g., around \(0.04\)),
and it can be mildly liberal at higher truncation levels (\(m\ge6\)) in several settings,
including some large-sample cases. This pattern is consistent with the known sensitivity of
orthogonal-series tests when the number of retained components approaches the limits of asymptotic validity.

\begin{table}[!ht]
\centering
\caption{Empirical Type~I error rates of the Stein-type test under the null
hypothesis \((X\sim p_N)\) at nominal level \(\alpha=0.05\).
Each entry reports the empirical rejection rate under \(H_0\) using
\emph{calibrated} critical values (Calibr.) and \emph{theoretical}
\(\chi^2\) critical values (Theoret.).
Calibrated critical values are estimated from \(50{,}000\) Monte Carlo draws under \(H_0\),
and Type~I error rates are evaluated on an independent \(20{,}000\) draws.}

\label{tab:type1_error}
\setlength{\tabcolsep}{10pt}
\renewcommand{\arraystretch}{1.1}
\begin{tabular}{cc cc cc cc cc}
\toprule
 & & \multicolumn{8}{c}{Empirical size under \(H_0\)} \\
\cmidrule(lr){3-10}
\(N\) & \(n\) & \multicolumn{2}{c}{\(m=4\)} & \multicolumn{2}{c}{\(m=6\)} & \multicolumn{2}{c}{\(m=8\)} & \multicolumn{2}{c}{\(m=10\)} \\
\cmidrule(lr){3-4}\cmidrule(lr){5-6}\cmidrule(lr){7-8}\cmidrule(lr){9-10}
 & & Calibr. & Theoret. & Calibr. & Theoret. & Calibr. & Theoret. & Calibr. & Theoret. \\
\midrule
\multirow{4}{*}{5}
  & 10  & 0.049 & 0.049 & 0.051 & 0.059 & 0.051 & 0.060 & 0.051 & 0.060 \\
  & 50  & 0.048 & 0.048 & 0.050 & 0.052 & 0.051 & 0.057 & 0.049 & 0.062 \\
  & 100  & 0.049 & 0.049 & 0.048 & 0.049 & 0.049 & 0.054 & 0.049 & 0.057 \\
  & 500  & 0.050 & 0.050 & 0.048 & 0.048 & 0.048 & 0.049 & 0.051 & 0.052 \\
\midrule
\multirow{4}{*}{10}
  & 10  & 0.050 & 0.050 & 0.051 & 0.049 & 0.050 & 0.049 & 0.049 & 0.051 \\
  & 50  & 0.055 & 0.052 & 0.057 & 0.062 & 0.056 & 0.059 & 0.055 & 0.057 \\
  & 100  & 0.047 & 0.048 & 0.048 & 0.055 & 0.049 & 0.059 & 0.047 & 0.056 \\
  & 500  & 0.051 & 0.052 & 0.054 & 0.055 & 0.051 & 0.059 & 0.051 & 0.062 \\
\midrule
\multirow{4}{*}{20}
  & 10  & 0.051 & 0.040 & 0.051 & 0.039 & 0.050 & 0.041 & 0.051 & 0.039 \\
  & 50  & 0.052 & 0.049 & 0.052 & 0.052 & 0.053 & 0.049 & 0.053 & 0.050 \\
  & 100  & 0.052 & 0.051 & 0.051 & 0.054 & 0.051 & 0.048 & 0.051 & 0.047 \\
  & 500  & 0.050 & 0.052 & 0.049 & 0.057 & 0.051 & 0.057 & 0.050 & 0.053 \\
\bottomrule
\end{tabular}
\end{table}

\begin{table}[!htb]
\centering
\caption{Estimated rejection rates (power) of the Stein-type test at truncation levels
\(m\in\{4,6,8,10\}\) against the Gaussian alternative \(X\sim\mathcal{N}(0,1)/\sqrt{N}\),
reported using calibrated (Calibr.) and theoretical \(\chi^2\) (Theoret.) critical values.
The nominal significance level is \(\alpha=0.05\).
Calibrated critical values are obtained from \(50{,}000\) Monte Carlo draws under \(H_0\),
and power is evaluated using \(20{,}000\) draws under \(H_1\).
For small systems (\(N=5\)), the test attains high power at moderate sample sizes, whereas for larger systems
(\(N\ge 10\)) substantially larger \(n\) is required to achieve comparable power.}

\label{tab:rejection_rates_compact}

\setlength{\tabcolsep}{10pt}
\renewcommand{\arraystretch}{1.1}
\begin{tabular}{cc cc cc cc cc}
\toprule
 & & \multicolumn{8}{c}{Empirical power under \(H_1\)} \\
\cmidrule(lr){3-10}
\(N\) & \(n\) & \multicolumn{2}{c}{\(m=4\)} & \multicolumn{2}{c}{\(m=6\)} & \multicolumn{2}{c}{\(m=8\)} & \multicolumn{2}{c}{\(m=10\)} \\
\cmidrule(lr){3-4}\cmidrule(lr){5-6}\cmidrule(lr){7-8}\cmidrule(lr){9-10}
 & & Calibr. & Theoret. & Calibr. & Theoret. & Calibr. & Theoret. & Calibr. & Theoret. \\
\midrule
\multirow{6}{*}{5}
  & 20  & 0.432 & 0.432 & 0.474 & 0.477 & 0.479 & 0.488 & 0.479 & 0.489 \\
  & 50  & 0.685 & 0.683 & 0.747 & 0.748 & 0.768 & 0.772 & 0.774 & 0.780 \\
  & 80  & 0.830 & 0.828 & 0.877 & 0.877 & 0.895 & 0.896 & 0.900 & 0.902 \\
  & 100  & 0.886 & 0.885 & 0.925 & 0.926 & 0.937 & 0.938 & 0.941 & 0.943 \\
  & 200  & 0.986 & 0.986 & 0.993 & 0.993 & 0.995 & 0.995 & 0.996 & 0.996 \\
  & 250  & 0.995 & 0.995 & 0.999 & 0.999 & 0.999 & 0.999 & 0.999 & 0.999 \\
\midrule
\multirow{5}{*}{10}
  & 50  & 0.317 & 0.313 & 0.331 & 0.337 & 0.328 & 0.334 & 0.325 & 0.329 \\
  & 100  & 0.434 & 0.436 & 0.467 & 0.477 & 0.466 & 0.482 & 0.463 & 0.474 \\
  & 150  & 0.558 & 0.557 & 0.589 & 0.597 & 0.587 & 0.602 & 0.584 & 0.595 \\
  & 200  & 0.638 & 0.641 & 0.673 & 0.681 & 0.673 & 0.687 & 0.670 & 0.683 \\
  & 500  & 0.914 & 0.915 & 0.927 & 0.928 & 0.925 & 0.930 & 0.921 & 0.927 \\
\midrule
\multirow{5}{*}{20}
  & 50  & 0.142 & 0.139 & 0.140 & 0.140 & 0.139 & 0.134 & 0.141 & 0.137 \\
  & 100  & 0.179 & 0.178 & 0.181 & 0.185 & 0.180 & 0.176 & 0.183 & 0.178 \\
  & 150  & 0.218 & 0.220 & 0.221 & 0.232 & 0.218 & 0.221 & 0.218 & 0.219 \\
  & 200  & 0.257 & 0.256 & 0.254 & 0.264 & 0.252 & 0.257 & 0.251 & 0.251 \\
  & 500  & 0.427 & 0.430 & 0.421 & 0.439 & 0.409 & 0.428 & 0.406 & 0.414 \\
\bottomrule
\end{tabular}
\end{table}

\subsection{Power Analysis}
Table~\ref{tab:rejection_rates_compact} reports estimated rejection rates (power) of the Stein-type test. The complete set of rejection rates over the full simulation grid
(all tested $N,n,m$ values, with both theoretical and calibrated cutoffs)
is reported in Appendix~\ref{app:full_power_table}.

\paragraph{Small systems (\(N=5\)).}
Against the Gaussian alternative, we take $X\sim\mathcal{N}(0,1)$ and apply the test to the rescaled variable $Y=X/\sqrt{N}$ (equivalently, $Y\sim\mathcal{N}(0,1/N)$).
The Stein-type test gains power rapidly as \(n\) increases.
At \(n=100\), the power is \(0.886\) for \(m=4\), while it exceeds \(0.9\) for \(m\ge 6\)
(\(0.925\) at \(m=6\), \(0.937\) at \(m=8\), and \(0.941\) at \(m=10\)).
By \(n=200\), power is already above \(0.98\) for all considered truncation levels,
and it is essentially one at \(n=250\).

\paragraph{Effect of the truncation level (\(m\)).}
For \(N=5\), increasing the truncation from \(m=4\) to \(m=6\) yields the largest improvement,
whereas further increases to \(m=8\) or \(m=10\) provide only marginal gains,
indicating saturation beyond \(m=6\) in this regime.

\paragraph{Larger systems (\(N=10\) and \(N=20\)).}
As \(N\) increases, the null \(p_N\) becomes closer to Gaussian, making discrimination more difficult at moderate \(n\).
For \(N=10\), power is around \(0.43\)--\(0.48\) at \(n=100\) and rises to about \(0.91\)--\(0.93\) by \(n=500\).
For \(N=20\), power remains low at \(n\le 200\) (\(\approx 0.13\)--\(0.27\)) and is still below \(0.5\) at \(n=500\)
(\(\approx 0.41\)--\(0.44\)), indicating that substantially larger sample sizes are required in this regime.

\begin{figure}[!h]
\centering
\includegraphics[width=0.90\textwidth]{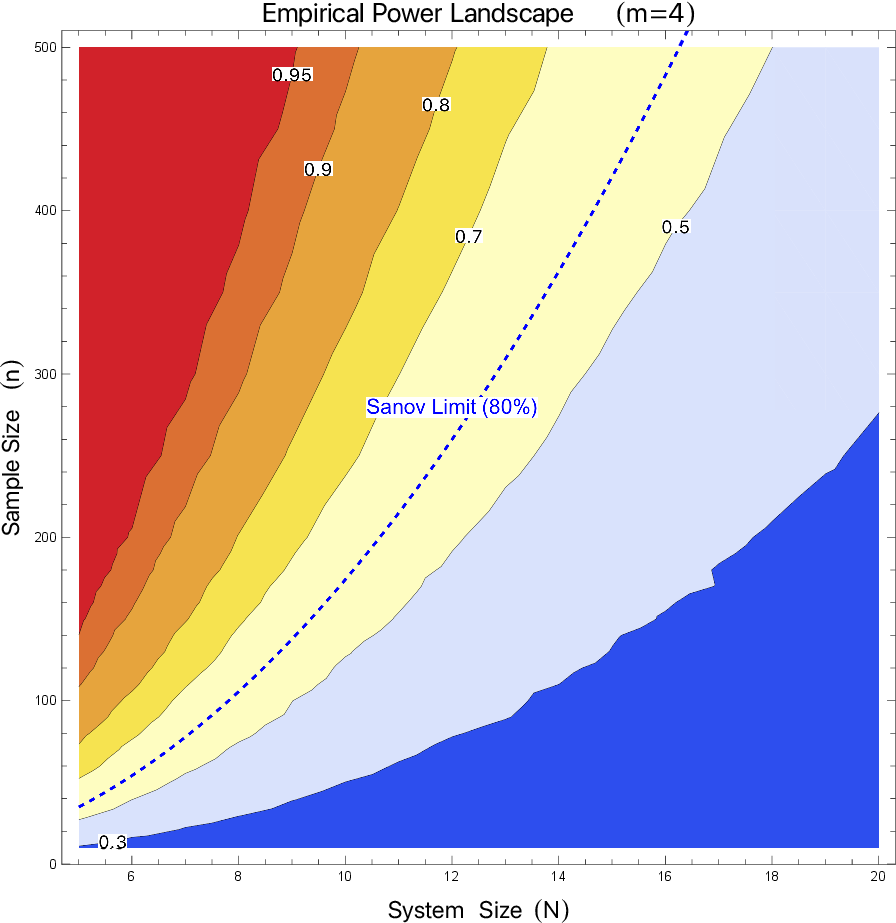}
\caption{
Empirical detection power (colour gradient) plotted together with the Sanov rate prediction (dashed curve).
Empirical results are obtained using the Stein-type test with truncation level \(m=4\) and the theoretical \(\chi^2\) critical values at \(\alpha=0.05\);
on our simulation grid, the corresponding calibrated cutoff yields nearly indistinguishable power.
The blue dashed curve indicates the theoretical \(80\%\)-power boundary derived from the Sanov exponent
\(\exp\{-n D_{\mathrm{KL}}(p_N\|p_\infty)\}\).
The empirical contours follow the same overall trend as the bound; the remaining gap for larger systems reflects the inherent trade-off
between extracting higher-order information and maintaining robust Type~I error control.
}
\label{fig:power_contour}
\end{figure}

\begin{figure}[htbp]
    \centering
    \includegraphics[width=0.95\textwidth]{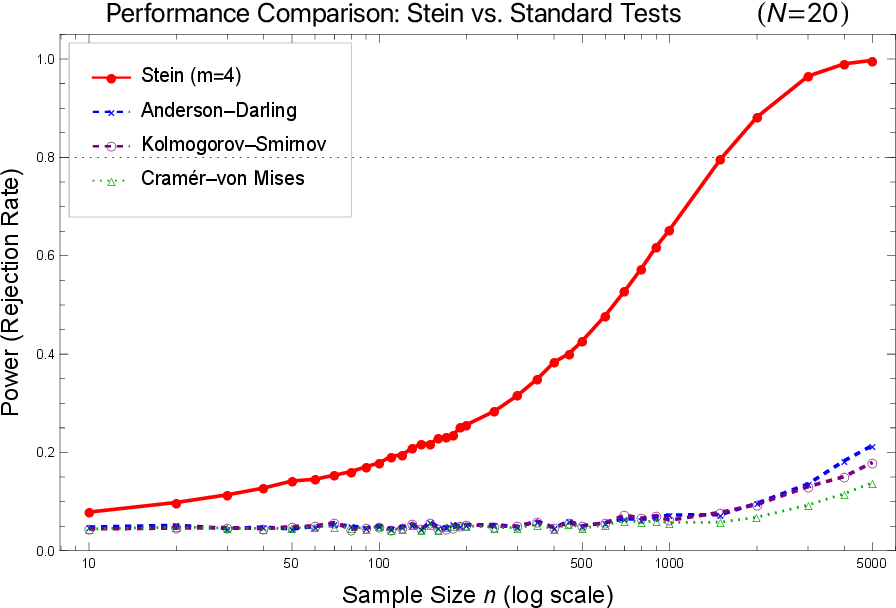}
    
    \caption{Power comparison between the proposed Stein-type test (\(m=4\)) and standard omnibus goodness-of-fit tests
    (Anderson--Darling, Kolmogorov--Smirnov, Cram\'er--von Mises)~\citep{stephens1974edf} for a system with \(N=20\).
    To ensure a fair comparison at the nominal level \(\alpha=0.05\), all curves report \emph{calibrated} power, i.e.,
    rejection rates computed using Monte Carlo calibrated critical values under \(H_0\).
    The horizontal axis shows the sample size \(n\) on a logarithmic scale.
    Over the displayed range (\(n\le 5000\)), the Stein-type test exhibits consistently higher power.}
    \label{fig:comparison_standard_tests}
\end{figure}

\begin{table}[htbp]
\centering
\caption{Empirical power of the Stein-type test for \(N=20\) over large sample sizes \(n\in[600,5000]\) at the nominal level \(\alpha=0.05\).
Each entry reports the rejection rate under \(H_1\) using \emph{calibrated} (Calibr.) and \emph{theoretical} \(\chi^2\) (Theoret.) critical values.
Calibrated critical values are estimated from \(50{,}000\) Monte Carlo draws under \(H_0\), and power is evaluated using \(20{,}000\) draws under the Gaussian alternative \(X\sim\mathcal{N}(0,1)/\sqrt{N}\).
As \(p_{20}\) is close to Gaussian, substantially larger samples are required to achieve high power than in small-\(N\) regimes.}

\label{tab:large_sample_N20}
\setlength{\tabcolsep}{10pt}
\renewcommand{\arraystretch}{1.15}
\begin{tabular}{cc cc cc cc cc}
\toprule
 & & \multicolumn{8}{c}{Empirical power under \(H_1\)} \\
\cmidrule(lr){3-10}
\(N\) & \(n\) & \multicolumn{2}{c}{\(m=4\)} & \multicolumn{2}{c}{\(m=6\)} & \multicolumn{2}{c}{\(m=8\)} & \multicolumn{2}{c}{\(m=10\)} \\
\cmidrule(lr){3-4}\cmidrule(lr){5-6}\cmidrule(lr){7-8}\cmidrule(lr){9-10}
 & & Calibr. & Theoret. & Calibr. & Theoret. & Calibr. & Theoret. & Calibr. & Theoret. \\
\midrule
\multirow{10}{*}{20}
  & 600  & 0.478 & 0.475 & 0.470 & 0.481 & 0.453 & 0.469 & 0.445 & 0.451 \\
  & 700  & 0.528 & 0.531 & 0.527 & 0.535 & 0.507 & 0.521 & 0.498 & 0.503 \\
  & 800  & 0.574 & 0.572 & 0.563 & 0.574 & 0.541 & 0.560 & 0.530 & 0.541 \\
  & 900  & 0.619 & 0.615 & 0.609 & 0.613 & 0.589 & 0.599 & 0.577 & 0.581 \\
  & 1000  & 0.653 & 0.647 & 0.637 & 0.643 & 0.618 & 0.631 & 0.608 & 0.610 \\
  & 1500  & 0.797 & 0.798 & 0.782 & 0.785 & 0.753 & 0.770 & 0.741 & 0.751 \\
  & 2000  & 0.882 & 0.883 & 0.871 & 0.874 & 0.847 & 0.858 & 0.835 & 0.842 \\
  & 3000  & 0.965 & 0.965 & 0.957 & 0.957 & 0.945 & 0.950 & 0.934 & 0.942 \\
  & 4000  & 0.990 & 0.990 & 0.986 & 0.987 & 0.979 & 0.982 & 0.975 & 0.978 \\
  & 5000  & 0.997 & 0.997 & 0.996 & 0.996 & 0.994 & 0.994 & 0.992 & 0.993 \\
\bottomrule
\end{tabular}
\end{table}

\subsection{Comparison with Sanov Bounds}
By the Chernoff--Stein lemma (often referred to as Stein's lemma), the optimal Type~II error exponent for testing
\(H_0:p_N\) against \(H_1:p_\infty\) at fixed Type~I level \(\alpha\in(0,1)\) satisfies
\[
\beta_n^{\ast}(\alpha)=\exp\{-nD_{\mathrm{KL}}(p_N\|p_\infty)+o(n)\}.
\]
This rate can be recovered from Sanov's theorem by analysing the large deviations of the empirical measure under the Neyman--Pearson test.
The closed-form expression used to evaluate $D_{\mathrm{KL}}(p_N\|p_\infty)$
is derived in Appendix~\ref{sec:derivation_DKL}.

For \(N=5\), we have \(D_{\mathrm{KL}}\simeq 0.0462\), so the large-deviation benchmark gives
\(\beta_n^{\ast}(\alpha)\approx \exp(-4.62)\approx 9.9\times 10^{-3}\) at \(n=100\),
corresponding to an exponent-only power proxy \(1-\exp(-nD_{\mathrm{KL}})\approx 0.99\) (See Table~\ref{tab:sanov_limit_exact}).
In comparison, our empirical power at \(N=5,n=100\) ranges from \(0.879\) at \(m=4\) to
\(0.920\) at \(m=6\) (and \(0.936\), \(0.941\) at \(m=8,10\)), indicating that finite-sample constants remain non-negligible.

For \(N=20\), the divergence is much smaller (\(D_{\mathrm{KL}}\simeq 0.00208\)), and even at \(n=200\) the benchmark proxy is only
\(1-\exp(-0.416)\approx 0.34\) (See Table~\ref{tab:sanov_limit_exact}), consistent with the low empirical power observed at moderate sample sizes
(e.g., \(\approx 0.25\)--\(0.26\) around \(n=200\) in Table~\ref{tab:rejection_rates_compact}).
This confirms that as \(N\) grows, \(p_N\) becomes increasingly close to Gaussian, and substantially larger \(n\) is required to detect the residual non-Gaussian signal.

Figure~\ref{fig:power_contour} summarises the empirical detection power of the proposed Stein-type test over the \((N,n)\) grid
(colour gradient) and overlays the large-deviation benchmark suggested by the Sanov/Chernoff--Stein exponent.
In particular, the dashed curve marks the nominal \(80\%\)-power boundary obtained from the approximation
\(\beta_n \approx \exp\{-n D_{\mathrm{KL}}(p_N\|p_\infty)\}\), i.e., the locus where
\(1-\exp\{-n D_{\mathrm{KL}}(p_N\|p_\infty)\}=0.8\).
The empirical contours follow the same qualitative trend: smaller systems (smaller \(N\)) are detectable at moderate \(n\),
while larger systems require substantially more samples as the finite-\(N\) distribution approaches the Gaussian limit.
Residual gaps between the empirical power and the Sanov-based boundary are consistent with finite-sample effects and with the practical
choice of truncation \(m=4\) combined with asymptotic \(\chi^2\) critical values, which prioritise stable Type~I control over extracting
very high-order features.

For completeness, a short derivation of the continuum rate functional
from the multinomial counting argument is given in Appendix~\ref{sec:sanov_functional}.

\begin{table}[htbp]
\centering
\caption{Sanov limit for the test power, \(1 - \exp[-n D_{\mathrm{KL}}(p_{N}\,\|\,p_{\infty})]\), using the exact closed-form divergence.  
As \(N\) grows the divergence shrinks, so a much larger sample size \(n\) is needed to reach the same power level.}

\label{tab:sanov_limit_exact}
\setlength{\tabcolsep}{3.5pt}
\begin{tabular}{c ccccccccc}
\toprule
 & \multicolumn{9}{c}{Sample Size (\(n\))} \\
\cmidrule(lr){2-10}
\textbf{\(N\)} & \textbf{10} & \textbf{50} & \textbf{100} & \textbf{200} & \textbf{400} & \textbf{600} & \textbf{800} & \textbf{1000} & \textbf{2000} \\
\midrule
\textbf{4}  & 0.555 & 0.983 & 1.000 & 1.000 & 1.000 & 1.000 & 1.000 & 1.000 & 1.000 \\
\textbf{5}  & 0.370 & 0.901 & 0.990 & 1.000 & 1.000 & 1.000 & 1.000 & 1.000 & 1.000 \\
\textbf{6}  & 0.257 & 0.774 & 0.949 & 0.997 & 1.000 & 1.000 & 1.000 & 1.000 & 1.000 \\
\textbf{8}  & 0.141 & 0.533 & 0.782 & 0.953 & 0.998 & 1.000 & 1.000 & 1.000 & 1.000 \\
\textbf{10} & 0.088 & 0.370 & 0.603 & 0.842 & 0.975 & 0.996 & 0.999 & 1.000 & 1.000 \\
\textbf{15} & 0.038 & 0.174 & 0.318 & 0.534 & 0.783 & 0.899 & 0.953 & 0.978 & 1.000 \\
\textbf{20} & 0.021 & 0.099 & 0.188 & 0.340 & 0.564 & 0.712 & 0.810 & 0.875 & 0.984 \\
\bottomrule
\end{tabular}
\end{table}

\subsection{Comparison to Standard Tests}

Figure~\ref{fig:comparison_standard_tests} provides a direct head-to-head comparison at \(N=20\)
against standard omnibus goodness-of-fit procedures (Anderson--Darling, Kolmogorov--Smirnov, and Cram\'er--von Mises).
All curves report \emph{calibrated} power at the nominal level \(\alpha=0.05\), meaning that critical values are obtained by Monte Carlo
under \(H_0\) so that each test attains the same Type~I error in finite samples.
Over the displayed range (\(n\le 5000\), with \(n\) on a logarithmic axis), the proposed Stein-type statistic with \(m=4\) achieves
consistently higher power, indicating that aligning the test direction with the finite-\(N\) Stein/Jacobi structure yields greater sensitivity
to the specific non-Gaussian deviations present in the microcanonical model.

\subsection{Practical Guidance}

\begin{itemize}
  \item \textbf{Truncation order.}
        We recommend \(m=4\) as the default choice, as it provides stable size control and competitive power across regimes.
        Using \(m=6\) can yield a noticeable power gain for small systems (e.g., \(N=5\)) at moderate \(n\),
        while larger truncation levels (\(m\ge 8\)) typically offer only marginal additional power.

  \item \textbf{Recommended sample sizes for 80\% power (Gaussian alternative).}
        Based on the reported simulation grid,
        \begin{itemize}
          \item \(\boldsymbol{N\approx 5}\): \(n\approx 80\) is sufficient (already above \(0.8\) at \(m=4\), and higher for \(m\ge 6\)).
          \item \(\boldsymbol{N\approx 10}\): \(n\approx 400\) (order of a few hundred) is required to reach \(\approx 80\%\) power,
                with \(n=500\) yielding power above \(0.9\).
          \item \(\boldsymbol{N=20}\): substantially larger samples are needed; power remains below \(0.5\) at \(n=500\),
                and reaching \(80\%\) requires \(n\) on the order of \(10^3\) (roughly \(n\approx 1{,}500\)--\(2{,}000\) depending on \(m\)).
        \end{itemize}
\end{itemize}


\section{Discussion}\label{sec:discussion}

\subsection{Optimal Truncation Level}

Across all $N$, the power curve generally flattens once $m \approx 6$.
Higher modes often add estimation variance (more degrees of freedom) than signal unless $n \gtrsim 500$.
We recommend $m=4\text{--}6$ for general applications.

\subsection{Outlook}

The one-dimensional treatment generalises to higher space dimensions by replacing Jacobi polynomials with Gegenbauer polynomials (and, more generally, spherical harmonics on the unit sphere).
The same numerical patterns---rapid saturation in $m$ and power governed by the Kullback--Leibler gap---are expected to hold.


\section{Concluding Remarks}

The exact finite-\(N\) velocity distribution, previously obtained from microcanonical geometry and recognised as the unique maximiser of Havrda--Charv\'at entropy under an energy constraint, served as our
starting point. By evaluating the likelihood ratio in a typical state, we recovered Sanov's large-deviation rate function (the Kullback--Leibler divergence) without Stirling or thermodynamic approximations.

Building on this result, we constructed a Stein operator that characterises the finite-\(N\) law and reduces to the classical Ornstein--Uhlenbeck form as \(N\to\infty\). The associated Jacobi polynomials yield an orthogonal test statistic that is simple to compute and has an asymptotic \(\chi^{2}\) limit (with degrees of freedom determined by the truncation level). Monte Carlo studies verified that the test controls Type I error and attains near-optimal power whenever the finite-size corrections are not masked by the thermodynamic limit.

The framework introduced here is ready for direct use in kinetic settings where the microcanonical finite-\(N\) law provides a principled non-Gaussian baseline and finite-size effects are non-negligible. Two lines of work now suggest themselves. First, the one-dimensional theory can be lifted to higher dimensions by replacing Jacobi polynomials with Gegenbauer polynomials (or, more generally, spherical bases on the unit sphere). Second, an adaptive scheme that estimates the effective particle number \(N\) from the data and then plugs this value into the Stein test would link statistical inference even more tightly to the underlying physics.

\backmatter

\section*{Statements and Declarations}

This work was partially supported by the KIST Institutional Program.\\
The authors have no relevant financial or non-financial interests to disclose.\\
JWS is the sole author and was responsible for the study conception and design, analysis, drafting and revising the manuscript, and approving the final version.

\begin{appendices}

\section{Full Simulation Results for Empirical Power Table}
\label{app:full_power_table}

Table~\ref{tab:full_simulation_results} reports empirical power (rejection rates under \(H_1\)) of the targeted Stein-type test for all combinations of particle number \(N\), sample size \(n\), and truncation level \(m\in\{4,6,8,10\}\) (even modes only). Each entry reports power using calibrated (Calibr.) and theoretical \(\chi^2\) (Theoret.) critical values at \(\alpha=0.05\). Calibrated critical values are estimated from 50,000 Monte Carlo draws under \(H_0\), and power is evaluated using 20,000 draws under \(H_1\).

\setlength{\tabcolsep}{8pt}
\renewcommand{\arraystretch}{1.10}
\begin{longtable}{c c cc cc cc cc}
\caption{Complete empirical power of the targeted Stein-type test (Calibr./Theoret.).}
\label{tab:full_simulation_results} \\
\toprule
\multirow{3}{*}{$N$} & \multirow{3}{*}{$n$} & \multicolumn{8}{c}{Power under $H_1$ (Calibr./Theoret.)} \\
\cmidrule(lr){3-10}
 & & \multicolumn{2}{c}{$m=4$} & \multicolumn{2}{c}{$m=6$} & \multicolumn{2}{c}{$m=8$} & \multicolumn{2}{c}{$m=10$} \\
\cmidrule(lr){3-4}\cmidrule(lr){5-6}\cmidrule(lr){7-8}\cmidrule(lr){9-10}
 & & Calibr. & Theoret. & Calibr. & Theoret. & Calibr. & Theoret. & Calibr. & Theoret. \\
\midrule
\endfirsthead

\multicolumn{10}{c}{{\bfseries \tablename\ \thetable{} -- continued from previous page}} \\
\toprule
\multirow{3}{*}{$N$} & \multirow{3}{*}{$n$} & \multicolumn{8}{c}{Power under $H_1$ (Calibr./Theoret.)} \\
\cmidrule(lr){3-10}
 & & \multicolumn{2}{c}{$m=4$} & \multicolumn{2}{c}{$m=6$} & \multicolumn{2}{c}{$m=8$} & \multicolumn{2}{c}{$m=10$} \\
\cmidrule(lr){3-4}\cmidrule(lr){5-6}\cmidrule(lr){7-8}\cmidrule(lr){9-10}
 & & Calibr. & Theoret. & Calibr. & Theoret. & Calibr. & Theoret. & Calibr. & Theoret. \\
\midrule
\endhead

\midrule
\multicolumn{10}{r}{{Continued on next page}} \\
\bottomrule
\endfoot

\bottomrule
\endlastfoot

\multirow[t]{26}{*}{5} & 10 & 0.287 & 0.287 & 0.302 & 0.309 & 0.302 & 0.308 & 0.299 & 0.306 \\
 & 20 & 0.432 & 0.432 & 0.474 & 0.477 & 0.479 & 0.488 & 0.479 & 0.489 \\
 & 30 & 0.532 & 0.529 & 0.588 & 0.591 & 0.600 & 0.608 & 0.602 & 0.610 \\
 & 40 & 0.614 & 0.613 & 0.678 & 0.680 & 0.697 & 0.701 & 0.701 & 0.707 \\
 & 50 & 0.685 & 0.683 & 0.747 & 0.748 & 0.768 & 0.772 & 0.774 & 0.780 \\
 & 60 & 0.739 & 0.739 & 0.800 & 0.801 & 0.821 & 0.824 & 0.826 & 0.831 \\
 & 70 & 0.790 & 0.791 & 0.845 & 0.847 & 0.863 & 0.866 & 0.869 & 0.873 \\
 & 80 & 0.830 & 0.828 & 0.877 & 0.877 & 0.895 & 0.896 & 0.900 & 0.902 \\
 & 90 & 0.865 & 0.864 & 0.906 & 0.906 & 0.921 & 0.922 & 0.926 & 0.928 \\
 & 100 & 0.886 & 0.885 & 0.925 & 0.926 & 0.937 & 0.938 & 0.941 & 0.943 \\
 & 110 & 0.910 & 0.909 & 0.942 & 0.941 & 0.952 & 0.953 & 0.956 & 0.957 \\
 & 120 & 0.926 & 0.925 & 0.956 & 0.956 & 0.965 & 0.965 & 0.968 & 0.968 \\
 & 130 & 0.938 & 0.937 & 0.962 & 0.962 & 0.971 & 0.972 & 0.975 & 0.975 \\
 & 140 & 0.949 & 0.950 & 0.971 & 0.971 & 0.977 & 0.978 & 0.979 & 0.980 \\
 & 150 & 0.962 & 0.961 & 0.979 & 0.979 & 0.983 & 0.983 & 0.986 & 0.986 \\
 & 160 & 0.967 & 0.968 & 0.982 & 0.982 & 0.987 & 0.987 & 0.989 & 0.989 \\
 & 170 & 0.972 & 0.972 & 0.985 & 0.985 & 0.989 & 0.989 & 0.991 & 0.991 \\
 & 180 & 0.977 & 0.977 & 0.987 & 0.988 & 0.991 & 0.991 & 0.992 & 0.992 \\
 & 190 & 0.983 & 0.983 & 0.992 & 0.992 & 0.994 & 0.994 & 0.995 & 0.996 \\
 & 200 & 0.986 & 0.986 & 0.993 & 0.993 & 0.995 & 0.995 & 0.996 & 0.996 \\
 & 250 & 0.995 & 0.995 & 0.999 & 0.999 & 0.999 & 0.999 & 0.999 & 0.999 \\
 & 300 & 0.999 & 0.999 & 0.999 & 0.999 & 1.000 & 1.000 & 1.000 & 1.000 \\
 & 350 & 1.000 & 1.000 & 1.000 & 1.000 & 1.000 & 1.000 & 1.000 & 1.000 \\
 & 400 & 1.000 & 1.000 & 1.000 & 1.000 & 1.000 & 1.000 & 1.000 & 1.000 \\
 & 450 & 1.000 & 1.000 & 1.000 & 1.000 & 1.000 & 1.000 & 1.000 & 1.000 \\
 & 500 & 1.000 & 1.000 & 1.000 & 1.000 & 1.000 & 1.000 & 1.000 & 1.000 \\
\midrule
\multirow[t]{26}{*}{6} & 10 & 0.227 & 0.229 & 0.236 & 0.244 & 0.235 & 0.242 & 0.234 & 0.239 \\
 & 20 & 0.339 & 0.339 & 0.366 & 0.373 & 0.366 & 0.376 & 0.363 & 0.373 \\
 & 30 & 0.436 & 0.437 & 0.476 & 0.483 & 0.479 & 0.488 & 0.477 & 0.489 \\
 & 40 & 0.509 & 0.505 & 0.558 & 0.562 & 0.563 & 0.574 & 0.562 & 0.574 \\
 & 50 & 0.569 & 0.571 & 0.623 & 0.627 & 0.632 & 0.642 & 0.633 & 0.644 \\
 & 60 & 0.625 & 0.623 & 0.679 & 0.680 & 0.691 & 0.697 & 0.692 & 0.702 \\
 & 70 & 0.672 & 0.675 & 0.728 & 0.730 & 0.742 & 0.750 & 0.743 & 0.753 \\
 & 80 & 0.712 & 0.711 & 0.768 & 0.770 & 0.782 & 0.788 & 0.785 & 0.792 \\
 & 90 & 0.748 & 0.748 & 0.801 & 0.802 & 0.818 & 0.823 & 0.820 & 0.826 \\
 & 100 & 0.783 & 0.782 & 0.830 & 0.832 & 0.846 & 0.850 & 0.848 & 0.854 \\
 & 110 & 0.806 & 0.804 & 0.852 & 0.853 & 0.867 & 0.869 & 0.870 & 0.875 \\
 & 120 & 0.828 & 0.828 & 0.877 & 0.877 & 0.890 & 0.893 & 0.892 & 0.897 \\
 & 130 & 0.859 & 0.859 & 0.897 & 0.898 & 0.907 & 0.909 & 0.911 & 0.914 \\
 & 140 & 0.876 & 0.875 & 0.909 & 0.910 & 0.922 & 0.924 & 0.925 & 0.928 \\
 & 150 & 0.891 & 0.891 & 0.925 & 0.926 & 0.935 & 0.936 & 0.938 & 0.940 \\
 & 160 & 0.910 & 0.909 & 0.938 & 0.938 & 0.947 & 0.948 & 0.947 & 0.950 \\
 & 170 & 0.919 & 0.918 & 0.944 & 0.945 & 0.954 & 0.955 & 0.956 & 0.957 \\
 & 180 & 0.929 & 0.929 & 0.953 & 0.954 & 0.961 & 0.962 & 0.962 & 0.964 \\
 & 190 & 0.941 & 0.941 & 0.963 & 0.964 & 0.970 & 0.970 & 0.971 & 0.973 \\
 & 200 & 0.947 & 0.946 & 0.968 & 0.968 & 0.972 & 0.973 & 0.974 & 0.974 \\
 & 250 & 0.975 & 0.975 & 0.985 & 0.985 & 0.989 & 0.989 & 0.989 & 0.989 \\
 & 300 & 0.989 & 0.989 & 0.994 & 0.994 & 0.995 & 0.995 & 0.995 & 0.996 \\
 & 350 & 0.994 & 0.994 & 0.997 & 0.997 & 0.998 & 0.998 & 0.998 & 0.998 \\
 & 400 & 0.997 & 0.997 & 0.999 & 0.999 & 0.999 & 0.999 & 0.999 & 0.999 \\
 & 450 & 0.999 & 0.999 & 1.000 & 1.000 & 1.000 & 1.000 & 1.000 & 1.000 \\
 & 500 & 1.000 & 1.000 & 1.000 & 1.000 & 1.000 & 1.000 & 1.000 & 1.000 \\
\midrule
\multirow[t]{26}{*}{7} & 10 & 0.186 & 0.186 & 0.189 & 0.194 & 0.185 & 0.190 & 0.184 & 0.191 \\
 & 20 & 0.283 & 0.283 & 0.299 & 0.308 & 0.297 & 0.306 & 0.297 & 0.304 \\
 & 30 & 0.348 & 0.350 & 0.377 & 0.387 & 0.377 & 0.388 & 0.374 & 0.386 \\
 & 40 & 0.427 & 0.425 & 0.458 & 0.467 & 0.459 & 0.473 & 0.456 & 0.470 \\
 & 50 & 0.469 & 0.469 & 0.515 & 0.523 & 0.519 & 0.532 & 0.519 & 0.531 \\
 & 60 & 0.526 & 0.527 & 0.575 & 0.580 & 0.581 & 0.594 & 0.581 & 0.594 \\
 & 70 & 0.567 & 0.567 & 0.614 & 0.619 & 0.622 & 0.633 & 0.620 & 0.633 \\
 & 80 & 0.605 & 0.603 & 0.658 & 0.664 & 0.666 & 0.678 & 0.666 & 0.680 \\
 & 90 & 0.644 & 0.645 & 0.699 & 0.702 & 0.708 & 0.717 & 0.707 & 0.720 \\
 & 100 & 0.682 & 0.682 & 0.731 & 0.735 & 0.742 & 0.750 & 0.741 & 0.752 \\
 & 110 & 0.707 & 0.709 & 0.758 & 0.762 & 0.769 & 0.778 & 0.768 & 0.778 \\
 & 120 & 0.737 & 0.735 & 0.781 & 0.784 & 0.790 & 0.798 & 0.788 & 0.798 \\
 & 130 & 0.759 & 0.758 & 0.804 & 0.806 & 0.813 & 0.820 & 0.813 & 0.823 \\
 & 140 & 0.787 & 0.785 & 0.828 & 0.830 & 0.838 & 0.843 & 0.837 & 0.846 \\
 & 150 & 0.802 & 0.802 & 0.844 & 0.847 & 0.855 & 0.860 & 0.856 & 0.863 \\
 & 160 & 0.823 & 0.821 & 0.864 & 0.864 & 0.872 & 0.876 & 0.874 & 0.879 \\
 & 170 & 0.836 & 0.837 & 0.875 & 0.877 & 0.885 & 0.891 & 0.886 & 0.892 \\
 & 180 & 0.852 & 0.851 & 0.885 & 0.886 & 0.894 & 0.898 & 0.895 & 0.901 \\
 & 190 & 0.871 & 0.869 & 0.901 & 0.902 & 0.907 & 0.911 & 0.908 & 0.913 \\
 & 200 & 0.881 & 0.880 & 0.911 & 0.911 & 0.918 & 0.921 & 0.919 & 0.923 \\
 & 250 & 0.931 & 0.930 & 0.950 & 0.951 & 0.957 & 0.958 & 0.956 & 0.959 \\
 & 300 & 0.958 & 0.957 & 0.972 & 0.972 & 0.975 & 0.976 & 0.975 & 0.977 \\
 & 350 & 0.976 & 0.976 & 0.985 & 0.985 & 0.986 & 0.987 & 0.987 & 0.988 \\
 & 400 & 0.987 & 0.987 & 0.992 & 0.992 & 0.993 & 0.993 & 0.993 & 0.993 \\
 & 450 & 0.992 & 0.993 & 0.996 & 0.996 & 0.996 & 0.996 & 0.996 & 0.996 \\
 & 500 & 0.996 & 0.996 & 0.997 & 0.997 & 0.998 & 0.998 & 0.998 & 0.998 \\
\midrule
\multirow[t]{26}{*}{8} & 10 & 0.165 & 0.165 & 0.166 & 0.169 & 0.163 & 0.164 & 0.162 & 0.167 \\
 & 20 & 0.243 & 0.244 & 0.250 & 0.260 & 0.248 & 0.255 & 0.247 & 0.254 \\
 & 30 & 0.304 & 0.305 & 0.322 & 0.333 & 0.321 & 0.330 & 0.318 & 0.327 \\
 & 40 & 0.357 & 0.360 & 0.382 & 0.393 & 0.379 & 0.394 & 0.378 & 0.388 \\
 & 50 & 0.400 & 0.401 & 0.427 & 0.439 & 0.426 & 0.443 & 0.426 & 0.440 \\
 & 60 & 0.443 & 0.442 & 0.478 & 0.486 & 0.477 & 0.491 & 0.475 & 0.487 \\
 & 70 & 0.487 & 0.484 & 0.523 & 0.531 & 0.524 & 0.538 & 0.522 & 0.534 \\
 & 80 & 0.518 & 0.521 & 0.558 & 0.568 & 0.561 & 0.577 & 0.560 & 0.574 \\
 & 90 & 0.553 & 0.552 & 0.596 & 0.603 & 0.597 & 0.613 & 0.597 & 0.612 \\
 & 100 & 0.583 & 0.582 & 0.622 & 0.630 & 0.628 & 0.644 & 0.625 & 0.642 \\
 & 110 & 0.614 & 0.614 & 0.655 & 0.662 & 0.659 & 0.671 & 0.659 & 0.671 \\
 & 120 & 0.641 & 0.643 & 0.690 & 0.695 & 0.694 & 0.708 & 0.692 & 0.706 \\
 & 130 & 0.666 & 0.665 & 0.710 & 0.714 & 0.716 & 0.727 & 0.712 & 0.725 \\
 & 140 & 0.691 & 0.693 & 0.731 & 0.736 & 0.737 & 0.749 & 0.737 & 0.749 \\
 & 150 & 0.711 & 0.712 & 0.752 & 0.756 & 0.754 & 0.765 & 0.754 & 0.767 \\
 & 160 & 0.731 & 0.731 & 0.772 & 0.778 & 0.777 & 0.787 & 0.774 & 0.788 \\
 & 170 & 0.755 & 0.752 & 0.793 & 0.796 & 0.797 & 0.805 & 0.796 & 0.806 \\
 & 180 & 0.766 & 0.769 & 0.805 & 0.810 & 0.811 & 0.821 & 0.810 & 0.822 \\
 & 190 & 0.791 & 0.789 & 0.825 & 0.826 & 0.832 & 0.838 & 0.829 & 0.837 \\
 & 200 & 0.803 & 0.802 & 0.838 & 0.841 & 0.845 & 0.852 & 0.845 & 0.853 \\
 & 250 & 0.869 & 0.869 & 0.896 & 0.897 & 0.901 & 0.905 & 0.898 & 0.905 \\
 & 300 & 0.909 & 0.909 & 0.931 & 0.932 & 0.933 & 0.936 & 0.931 & 0.937 \\
 & 350 & 0.939 & 0.938 & 0.955 & 0.955 & 0.957 & 0.959 & 0.955 & 0.958 \\
 & 400 & 0.958 & 0.958 & 0.971 & 0.971 & 0.974 & 0.975 & 0.973 & 0.974 \\
 & 450 & 0.973 & 0.972 & 0.981 & 0.981 & 0.983 & 0.984 & 0.983 & 0.984 \\
 & 500 & 0.982 & 0.982 & 0.989 & 0.989 & 0.989 & 0.990 & 0.988 & 0.990 \\
\midrule
\multirow[t]{26}{*}{9} & 10 & 0.147 & 0.146 & 0.149 & 0.148 & 0.148 & 0.147 & 0.150 & 0.152 \\
 & 20 & 0.214 & 0.212 & 0.215 & 0.221 & 0.214 & 0.216 & 0.213 & 0.217 \\
 & 30 & 0.263 & 0.264 & 0.273 & 0.283 & 0.274 & 0.281 & 0.271 & 0.277 \\
 & 40 & 0.299 & 0.298 & 0.318 & 0.327 & 0.313 & 0.322 & 0.312 & 0.318 \\
 & 50 & 0.341 & 0.341 & 0.362 & 0.374 & 0.361 & 0.373 & 0.358 & 0.368 \\
 & 60 & 0.381 & 0.379 & 0.405 & 0.415 & 0.402 & 0.413 & 0.400 & 0.408 \\
 & 70 & 0.420 & 0.416 & 0.444 & 0.457 & 0.442 & 0.457 & 0.442 & 0.452 \\
 & 80 & 0.442 & 0.443 & 0.476 & 0.486 & 0.473 & 0.488 & 0.472 & 0.485 \\
 & 90 & 0.471 & 0.472 & 0.505 & 0.516 & 0.503 & 0.520 & 0.502 & 0.516 \\
 & 100 & 0.508 & 0.503 & 0.538 & 0.547 & 0.539 & 0.554 & 0.536 & 0.551 \\
 & 110 & 0.531 & 0.532 & 0.571 & 0.580 & 0.572 & 0.588 & 0.570 & 0.582 \\
 & 120 & 0.562 & 0.562 & 0.600 & 0.607 & 0.601 & 0.614 & 0.597 & 0.609 \\
 & 130 & 0.578 & 0.578 & 0.621 & 0.626 & 0.621 & 0.634 & 0.618 & 0.631 \\
 & 140 & 0.606 & 0.604 & 0.641 & 0.650 & 0.642 & 0.657 & 0.638 & 0.652 \\
 & 150 & 0.628 & 0.630 & 0.667 & 0.673 & 0.668 & 0.683 & 0.666 & 0.679 \\
 & 160 & 0.650 & 0.649 & 0.686 & 0.692 & 0.690 & 0.702 & 0.687 & 0.701 \\
 & 170 & 0.668 & 0.669 & 0.707 & 0.711 & 0.707 & 0.719 & 0.704 & 0.718 \\
 & 180 & 0.687 & 0.686 & 0.723 & 0.728 & 0.723 & 0.734 & 0.719 & 0.731 \\
 & 190 & 0.701 & 0.703 & 0.737 & 0.741 & 0.738 & 0.748 & 0.736 & 0.748 \\
 & 200 & 0.721 & 0.722 & 0.754 & 0.761 & 0.753 & 0.768 & 0.753 & 0.765 \\
 & 250 & 0.791 & 0.791 & 0.823 & 0.826 & 0.824 & 0.834 & 0.820 & 0.832 \\
 & 300 & 0.842 & 0.841 & 0.868 & 0.871 & 0.868 & 0.877 & 0.865 & 0.874 \\
 & 350 & 0.886 & 0.886 & 0.905 & 0.905 & 0.908 & 0.912 & 0.905 & 0.912 \\
 & 400 & 0.915 & 0.915 & 0.931 & 0.932 & 0.932 & 0.936 & 0.930 & 0.935 \\
 & 450 & 0.940 & 0.939 & 0.951 & 0.951 & 0.952 & 0.954 & 0.950 & 0.954 \\
 & 500 & 0.958 & 0.957 & 0.966 & 0.967 & 0.966 & 0.969 & 0.965 & 0.968 \\
\midrule
\multirow[t]{26}{*}{10} & 10 & 0.125 & 0.125 & 0.127 & 0.125 & 0.125 & 0.124 & 0.126 & 0.128 \\
 & 20 & 0.188 & 0.189 & 0.190 & 0.196 & 0.186 & 0.188 & 0.188 & 0.192 \\
 & 30 & 0.233 & 0.234 & 0.240 & 0.247 & 0.237 & 0.242 & 0.235 & 0.241 \\
 & 40 & 0.263 & 0.263 & 0.276 & 0.286 & 0.275 & 0.279 & 0.273 & 0.278 \\
 & 50 & 0.317 & 0.313 & 0.331 & 0.337 & 0.328 & 0.334 & 0.325 & 0.329 \\
 & 60 & 0.332 & 0.334 & 0.350 & 0.363 & 0.349 & 0.361 & 0.348 & 0.357 \\
 & 70 & 0.366 & 0.364 & 0.384 & 0.395 & 0.381 & 0.392 & 0.380 & 0.387 \\
 & 80 & 0.384 & 0.384 & 0.406 & 0.420 & 0.403 & 0.419 & 0.399 & 0.413 \\
 & 90 & 0.421 & 0.420 & 0.444 & 0.457 & 0.441 & 0.458 & 0.439 & 0.452 \\
 & 100 & 0.434 & 0.436 & 0.467 & 0.477 & 0.466 & 0.482 & 0.463 & 0.474 \\
 & 110 & 0.467 & 0.467 & 0.495 & 0.506 & 0.492 & 0.506 & 0.489 & 0.501 \\
 & 120 & 0.492 & 0.489 & 0.523 & 0.530 & 0.521 & 0.532 & 0.518 & 0.526 \\
 & 130 & 0.517 & 0.514 & 0.543 & 0.552 & 0.541 & 0.556 & 0.542 & 0.552 \\
 & 140 & 0.529 & 0.529 & 0.557 & 0.567 & 0.554 & 0.570 & 0.551 & 0.565 \\
 & 150 & 0.558 & 0.557 & 0.589 & 0.597 & 0.587 & 0.602 & 0.584 & 0.595 \\
 & 160 & 0.572 & 0.575 & 0.605 & 0.615 & 0.602 & 0.619 & 0.598 & 0.614 \\
 & 170 & 0.591 & 0.593 & 0.618 & 0.631 & 0.616 & 0.635 & 0.612 & 0.630 \\
 & 180 & 0.610 & 0.610 & 0.640 & 0.649 & 0.637 & 0.654 & 0.634 & 0.649 \\
 & 190 & 0.630 & 0.627 & 0.660 & 0.664 & 0.657 & 0.667 & 0.654 & 0.665 \\
 & 200 & 0.638 & 0.641 & 0.673 & 0.681 & 0.673 & 0.687 & 0.670 & 0.683 \\
 & 250 & 0.716 & 0.715 & 0.743 & 0.749 & 0.742 & 0.753 & 0.736 & 0.750 \\
 & 300 & 0.773 & 0.773 & 0.796 & 0.800 & 0.796 & 0.807 & 0.791 & 0.804 \\
 & 350 & 0.823 & 0.820 & 0.840 & 0.842 & 0.838 & 0.847 & 0.831 & 0.841 \\
 & 400 & 0.857 & 0.857 & 0.875 & 0.877 & 0.873 & 0.880 & 0.869 & 0.877 \\
 & 450 & 0.890 & 0.889 & 0.906 & 0.907 & 0.902 & 0.908 & 0.898 & 0.906 \\
 & 500 & 0.914 & 0.915 & 0.927 & 0.928 & 0.925 & 0.930 & 0.921 & 0.927 \\
\midrule
\multirow[t]{26}{*}{11} & 10 & 0.120 & 0.117 & 0.117 & 0.114 & 0.118 & 0.115 & 0.118 & 0.118 \\
 & 20 & 0.167 & 0.168 & 0.170 & 0.174 & 0.167 & 0.168 & 0.170 & 0.173 \\
 & 30 & 0.208 & 0.208 & 0.213 & 0.218 & 0.208 & 0.211 & 0.209 & 0.214 \\
 & 40 & 0.233 & 0.235 & 0.241 & 0.250 & 0.237 & 0.243 & 0.239 & 0.243 \\
 & 50 & 0.273 & 0.273 & 0.280 & 0.290 & 0.275 & 0.284 & 0.274 & 0.280 \\
 & 60 & 0.292 & 0.291 & 0.303 & 0.316 & 0.301 & 0.310 & 0.297 & 0.305 \\
 & 70 & 0.323 & 0.325 & 0.336 & 0.351 & 0.332 & 0.345 & 0.334 & 0.341 \\
 & 80 & 0.344 & 0.345 & 0.360 & 0.372 & 0.358 & 0.369 & 0.354 & 0.361 \\
 & 90 & 0.369 & 0.369 & 0.387 & 0.398 & 0.380 & 0.392 & 0.377 & 0.385 \\
 & 100 & 0.395 & 0.395 & 0.416 & 0.426 & 0.411 & 0.423 & 0.407 & 0.415 \\
 & 110 & 0.414 & 0.414 & 0.434 & 0.446 & 0.432 & 0.445 & 0.429 & 0.437 \\
 & 120 & 0.435 & 0.436 & 0.461 & 0.472 & 0.458 & 0.471 & 0.454 & 0.465 \\
 & 130 & 0.456 & 0.455 & 0.486 & 0.495 & 0.481 & 0.494 & 0.479 & 0.485 \\
 & 140 & 0.464 & 0.466 & 0.495 & 0.506 & 0.490 & 0.507 & 0.486 & 0.498 \\
 & 150 & 0.484 & 0.483 & 0.509 & 0.522 & 0.504 & 0.521 & 0.500 & 0.512 \\
 & 160 & 0.511 & 0.511 & 0.539 & 0.548 & 0.534 & 0.548 & 0.529 & 0.540 \\
 & 170 & 0.528 & 0.526 & 0.554 & 0.563 & 0.548 & 0.565 & 0.544 & 0.557 \\
 & 180 & 0.540 & 0.541 & 0.561 & 0.574 & 0.560 & 0.576 & 0.557 & 0.569 \\
 & 190 & 0.555 & 0.556 & 0.582 & 0.591 & 0.577 & 0.593 & 0.575 & 0.587 \\
 & 200 & 0.572 & 0.572 & 0.601 & 0.610 & 0.596 & 0.612 & 0.594 & 0.607 \\
 & 250 & 0.641 & 0.640 & 0.665 & 0.675 & 0.663 & 0.677 & 0.656 & 0.671 \\
 & 300 & 0.699 & 0.700 & 0.725 & 0.733 & 0.719 & 0.735 & 0.712 & 0.727 \\
 & 350 & 0.755 & 0.755 & 0.774 & 0.778 & 0.768 & 0.782 & 0.761 & 0.774 \\
 & 400 & 0.800 & 0.800 & 0.817 & 0.822 & 0.812 & 0.824 & 0.806 & 0.819 \\
 & 450 & 0.834 & 0.835 & 0.852 & 0.853 & 0.843 & 0.851 & 0.836 & 0.847 \\
 & 500 & 0.866 & 0.865 & 0.880 & 0.883 & 0.872 & 0.880 & 0.867 & 0.876 \\
\midrule
\multirow[t]{26}{*}{12} & 10 & 0.110 & 0.107 & 0.108 & 0.102 & 0.110 & 0.106 & 0.110 & 0.108 \\
 & 20 & 0.152 & 0.151 & 0.152 & 0.152 & 0.148 & 0.147 & 0.150 & 0.151 \\
 & 30 & 0.184 & 0.184 & 0.190 & 0.194 & 0.186 & 0.186 & 0.186 & 0.189 \\
 & 40 & 0.211 & 0.213 & 0.216 & 0.225 & 0.214 & 0.217 & 0.213 & 0.217 \\
 & 50 & 0.235 & 0.235 & 0.246 & 0.253 & 0.242 & 0.246 & 0.243 & 0.244 \\
 & 60 & 0.262 & 0.263 & 0.270 & 0.283 & 0.269 & 0.277 & 0.268 & 0.273 \\
 & 70 & 0.289 & 0.286 & 0.302 & 0.309 & 0.296 & 0.305 & 0.294 & 0.299 \\
 & 80 & 0.307 & 0.305 & 0.317 & 0.328 & 0.314 & 0.323 & 0.316 & 0.321 \\
 & 90 & 0.328 & 0.327 & 0.342 & 0.354 & 0.339 & 0.349 & 0.338 & 0.343 \\
 & 100 & 0.354 & 0.353 & 0.369 & 0.380 & 0.365 & 0.373 & 0.365 & 0.369 \\
 & 110 & 0.371 & 0.370 & 0.390 & 0.401 & 0.385 & 0.397 & 0.385 & 0.392 \\
 & 120 & 0.388 & 0.387 & 0.407 & 0.417 & 0.405 & 0.416 & 0.400 & 0.407 \\
 & 130 & 0.401 & 0.398 & 0.416 & 0.428 & 0.411 & 0.425 & 0.410 & 0.418 \\
 & 140 & 0.420 & 0.420 & 0.439 & 0.453 & 0.437 & 0.451 & 0.435 & 0.444 \\
 & 150 & 0.435 & 0.432 & 0.454 & 0.464 & 0.451 & 0.466 & 0.454 & 0.460 \\
 & 160 & 0.451 & 0.450 & 0.468 & 0.482 & 0.465 & 0.480 & 0.462 & 0.472 \\
 & 170 & 0.469 & 0.472 & 0.491 & 0.505 & 0.487 & 0.504 & 0.484 & 0.496 \\
 & 180 & 0.477 & 0.477 & 0.503 & 0.514 & 0.497 & 0.511 & 0.494 & 0.503 \\
 & 190 & 0.496 & 0.496 & 0.520 & 0.529 & 0.513 & 0.527 & 0.510 & 0.517 \\
 & 200 & 0.516 & 0.515 & 0.540 & 0.549 & 0.533 & 0.548 & 0.531 & 0.539 \\
 & 250 & 0.582 & 0.580 & 0.606 & 0.612 & 0.596 & 0.612 & 0.596 & 0.604 \\
 & 300 & 0.633 & 0.637 & 0.655 & 0.664 & 0.651 & 0.666 & 0.645 & 0.658 \\
 & 350 & 0.684 & 0.686 & 0.704 & 0.712 & 0.700 & 0.714 & 0.694 & 0.706 \\
 & 400 & 0.737 & 0.735 & 0.749 & 0.753 & 0.745 & 0.754 & 0.738 & 0.748 \\
 & 450 & 0.776 & 0.776 & 0.790 & 0.793 & 0.782 & 0.794 & 0.775 & 0.786 \\
 & 500 & 0.803 & 0.803 & 0.817 & 0.822 & 0.808 & 0.821 & 0.803 & 0.813 \\
\midrule
\multirow[t]{26}{*}{13} & 10 & 0.105 & 0.101 & 0.104 & 0.097 & 0.106 & 0.101 & 0.105 & 0.100 \\
 & 20 & 0.141 & 0.141 & 0.141 & 0.141 & 0.140 & 0.137 & 0.142 & 0.143 \\
 & 30 & 0.169 & 0.170 & 0.172 & 0.175 & 0.168 & 0.168 & 0.171 & 0.172 \\
 & 40 & 0.196 & 0.196 & 0.199 & 0.204 & 0.198 & 0.199 & 0.198 & 0.201 \\
 & 50 & 0.218 & 0.218 & 0.223 & 0.231 & 0.221 & 0.224 & 0.221 & 0.223 \\
 & 60 & 0.237 & 0.237 & 0.245 & 0.254 & 0.244 & 0.247 & 0.240 & 0.243 \\
 & 70 & 0.263 & 0.262 & 0.272 & 0.278 & 0.269 & 0.271 & 0.268 & 0.270 \\
 & 80 & 0.275 & 0.273 & 0.281 & 0.294 & 0.277 & 0.288 & 0.275 & 0.285 \\
 & 90 & 0.288 & 0.288 & 0.297 & 0.309 & 0.297 & 0.303 & 0.296 & 0.299 \\
 & 100 & 0.319 & 0.317 & 0.332 & 0.343 & 0.331 & 0.339 & 0.331 & 0.334 \\
 & 110 & 0.321 & 0.322 & 0.337 & 0.352 & 0.338 & 0.347 & 0.337 & 0.341 \\
 & 120 & 0.346 & 0.348 & 0.360 & 0.375 & 0.352 & 0.367 & 0.351 & 0.360 \\
 & 130 & 0.363 & 0.358 & 0.375 & 0.384 & 0.374 & 0.382 & 0.370 & 0.375 \\
 & 140 & 0.376 & 0.376 & 0.393 & 0.405 & 0.390 & 0.400 & 0.387 & 0.393 \\
 & 150 & 0.397 & 0.396 & 0.414 & 0.425 & 0.407 & 0.419 & 0.404 & 0.412 \\
 & 160 & 0.407 & 0.407 & 0.428 & 0.438 & 0.420 & 0.432 & 0.416 & 0.423 \\
 & 170 & 0.418 & 0.413 & 0.436 & 0.444 & 0.431 & 0.442 & 0.427 & 0.430 \\
 & 180 & 0.428 & 0.429 & 0.447 & 0.461 & 0.443 & 0.459 & 0.441 & 0.451 \\
 & 190 & 0.445 & 0.443 & 0.465 & 0.474 & 0.459 & 0.470 & 0.457 & 0.463 \\
 & 200 & 0.462 & 0.462 & 0.476 & 0.489 & 0.475 & 0.487 & 0.473 & 0.479 \\
 & 250 & 0.515 & 0.515 & 0.534 & 0.545 & 0.527 & 0.542 & 0.520 & 0.531 \\
 & 300 & 0.577 & 0.575 & 0.594 & 0.604 & 0.582 & 0.601 & 0.580 & 0.593 \\
 & 350 & 0.629 & 0.629 & 0.643 & 0.650 & 0.631 & 0.646 & 0.625 & 0.638 \\
 & 400 & 0.667 & 0.670 & 0.688 & 0.694 & 0.678 & 0.691 & 0.672 & 0.682 \\
 & 450 & 0.708 & 0.712 & 0.721 & 0.731 & 0.712 & 0.728 & 0.708 & 0.719 \\
 & 500 & 0.750 & 0.750 & 0.756 & 0.763 & 0.744 & 0.760 & 0.739 & 0.749 \\
\midrule
\multirow[t]{26}{*}{14} & 10 & 0.099 & 0.093 & 0.097 & 0.089 & 0.099 & 0.093 & 0.099 & 0.093 \\
 & 20 & 0.135 & 0.132 & 0.133 & 0.128 & 0.132 & 0.126 & 0.132 & 0.130 \\
 & 30 & 0.159 & 0.158 & 0.158 & 0.159 & 0.157 & 0.155 & 0.157 & 0.158 \\
 & 40 & 0.180 & 0.179 & 0.181 & 0.185 & 0.178 & 0.177 & 0.181 & 0.181 \\
 & 50 & 0.201 & 0.200 & 0.202 & 0.209 & 0.201 & 0.203 & 0.203 & 0.204 \\
 & 60 & 0.218 & 0.219 & 0.222 & 0.233 & 0.220 & 0.223 & 0.222 & 0.224 \\
 & 70 & 0.237 & 0.237 & 0.243 & 0.253 & 0.241 & 0.246 & 0.241 & 0.243 \\
 & 80 & 0.255 & 0.255 & 0.259 & 0.270 & 0.259 & 0.265 & 0.257 & 0.261 \\
 & 90 & 0.263 & 0.265 & 0.273 & 0.283 & 0.270 & 0.277 & 0.269 & 0.273 \\
 & 100 & 0.284 & 0.283 & 0.291 & 0.300 & 0.288 & 0.294 & 0.287 & 0.290 \\
 & 110 & 0.293 & 0.289 & 0.310 & 0.316 & 0.302 & 0.309 & 0.302 & 0.302 \\
 & 120 & 0.312 & 0.316 & 0.327 & 0.339 & 0.325 & 0.333 & 0.324 & 0.328 \\
 & 130 & 0.329 & 0.325 & 0.340 & 0.351 & 0.335 & 0.343 & 0.335 & 0.336 \\
 & 140 & 0.340 & 0.339 & 0.353 & 0.364 & 0.347 & 0.356 & 0.344 & 0.349 \\
 & 150 & 0.351 & 0.351 & 0.364 & 0.377 & 0.361 & 0.373 & 0.359 & 0.364 \\
 & 160 & 0.361 & 0.359 & 0.374 & 0.387 & 0.370 & 0.384 & 0.369 & 0.376 \\
 & 170 & 0.383 & 0.387 & 0.397 & 0.411 & 0.392 & 0.406 & 0.391 & 0.399 \\
 & 180 & 0.399 & 0.397 & 0.411 & 0.421 & 0.404 & 0.416 & 0.403 & 0.408 \\
 & 190 & 0.397 & 0.398 & 0.412 & 0.424 & 0.410 & 0.420 & 0.404 & 0.411 \\
 & 200 & 0.413 & 0.415 & 0.429 & 0.442 & 0.423 & 0.438 & 0.418 & 0.428 \\
 & 250 & 0.477 & 0.476 & 0.489 & 0.501 & 0.480 & 0.497 & 0.478 & 0.487 \\
 & 300 & 0.523 & 0.524 & 0.543 & 0.551 & 0.530 & 0.545 & 0.527 & 0.534 \\
 & 350 & 0.575 & 0.573 & 0.588 & 0.598 & 0.579 & 0.593 & 0.573 & 0.583 \\
 & 400 & 0.621 & 0.619 & 0.631 & 0.638 & 0.618 & 0.636 & 0.611 & 0.623 \\
 & 450 & 0.652 & 0.650 & 0.662 & 0.666 & 0.646 & 0.660 & 0.642 & 0.649 \\
 & 500 & 0.687 & 0.686 & 0.695 & 0.704 & 0.682 & 0.699 & 0.677 & 0.688 \\
\midrule
\multirow[t]{26}{*}{15} & 10 & 0.095 & 0.085 & 0.091 & 0.081 & 0.092 & 0.085 & 0.091 & 0.083 \\
 & 20 & 0.129 & 0.127 & 0.125 & 0.124 & 0.125 & 0.122 & 0.127 & 0.125 \\
 & 30 & 0.144 & 0.144 & 0.146 & 0.145 & 0.144 & 0.142 & 0.145 & 0.145 \\
 & 40 & 0.160 & 0.161 & 0.163 & 0.166 & 0.162 & 0.161 & 0.164 & 0.164 \\
 & 50 & 0.180 & 0.180 & 0.182 & 0.186 & 0.180 & 0.179 & 0.179 & 0.180 \\
 & 60 & 0.203 & 0.203 & 0.206 & 0.211 & 0.203 & 0.204 & 0.201 & 0.202 \\
 & 70 & 0.217 & 0.216 & 0.217 & 0.226 & 0.217 & 0.218 & 0.218 & 0.219 \\
 & 80 & 0.232 & 0.230 & 0.234 & 0.243 & 0.233 & 0.236 & 0.234 & 0.235 \\
 & 90 & 0.247 & 0.246 & 0.255 & 0.263 & 0.252 & 0.254 & 0.250 & 0.250 \\
 & 100 & 0.267 & 0.265 & 0.271 & 0.281 & 0.266 & 0.271 & 0.265 & 0.267 \\
 & 110 & 0.270 & 0.270 & 0.270 & 0.285 & 0.268 & 0.275 & 0.267 & 0.271 \\
 & 120 & 0.289 & 0.287 & 0.294 & 0.307 & 0.293 & 0.299 & 0.292 & 0.293 \\
 & 130 & 0.304 & 0.302 & 0.310 & 0.319 & 0.308 & 0.312 & 0.304 & 0.304 \\
 & 140 & 0.311 & 0.313 & 0.321 & 0.335 & 0.318 & 0.327 & 0.314 & 0.319 \\
 & 150 & 0.322 & 0.324 & 0.333 & 0.346 & 0.330 & 0.339 & 0.328 & 0.330 \\
 & 160 & 0.333 & 0.333 & 0.345 & 0.357 & 0.340 & 0.350 & 0.339 & 0.341 \\
 & 170 & 0.349 & 0.346 & 0.359 & 0.368 & 0.358 & 0.363 & 0.356 & 0.355 \\
 & 180 & 0.353 & 0.353 & 0.364 & 0.378 & 0.360 & 0.370 & 0.358 & 0.363 \\
 & 190 & 0.362 & 0.365 & 0.375 & 0.388 & 0.371 & 0.383 & 0.366 & 0.373 \\
 & 200 & 0.381 & 0.378 & 0.389 & 0.403 & 0.384 & 0.395 & 0.380 & 0.385 \\
 & 250 & 0.427 & 0.425 & 0.435 & 0.447 & 0.431 & 0.444 & 0.425 & 0.431 \\
 & 300 & 0.475 & 0.473 & 0.488 & 0.498 & 0.479 & 0.491 & 0.476 & 0.481 \\
 & 350 & 0.524 & 0.520 & 0.533 & 0.541 & 0.521 & 0.535 & 0.517 & 0.524 \\
 & 400 & 0.569 & 0.564 & 0.575 & 0.581 & 0.561 & 0.574 & 0.560 & 0.564 \\
 & 450 & 0.601 & 0.603 & 0.609 & 0.619 & 0.596 & 0.613 & 0.589 & 0.601 \\
 & 500 & 0.629 & 0.631 & 0.642 & 0.652 & 0.633 & 0.647 & 0.626 & 0.636 \\
\midrule
\multirow[t]{26}{*}{16} & 10 & 0.090 & 0.083 & 0.090 & 0.079 & 0.091 & 0.082 & 0.090 & 0.079 \\
 & 20 & 0.113 & 0.112 & 0.113 & 0.110 & 0.113 & 0.108 & 0.115 & 0.111 \\
 & 30 & 0.142 & 0.139 & 0.142 & 0.139 & 0.140 & 0.134 & 0.144 & 0.140 \\
 & 40 & 0.151 & 0.153 & 0.151 & 0.156 & 0.152 & 0.152 & 0.153 & 0.156 \\
 & 50 & 0.172 & 0.171 & 0.174 & 0.177 & 0.172 & 0.170 & 0.173 & 0.172 \\
 & 60 & 0.182 & 0.181 & 0.187 & 0.188 & 0.185 & 0.182 & 0.183 & 0.181 \\
 & 70 & 0.206 & 0.203 & 0.210 & 0.216 & 0.205 & 0.205 & 0.206 & 0.207 \\
 & 80 & 0.207 & 0.207 & 0.213 & 0.220 & 0.213 & 0.214 & 0.212 & 0.212 \\
 & 90 & 0.226 & 0.224 & 0.231 & 0.238 & 0.229 & 0.230 & 0.227 & 0.228 \\
 & 100 & 0.243 & 0.244 & 0.250 & 0.258 & 0.247 & 0.249 & 0.247 & 0.246 \\
 & 110 & 0.252 & 0.252 & 0.259 & 0.268 & 0.254 & 0.258 & 0.254 & 0.255 \\
 & 120 & 0.261 & 0.262 & 0.269 & 0.281 & 0.269 & 0.273 & 0.268 & 0.270 \\
 & 130 & 0.281 & 0.278 & 0.290 & 0.298 & 0.283 & 0.288 & 0.282 & 0.284 \\
 & 140 & 0.283 & 0.281 & 0.290 & 0.302 & 0.282 & 0.291 & 0.282 & 0.285 \\
 & 150 & 0.288 & 0.290 & 0.298 & 0.310 & 0.294 & 0.304 & 0.294 & 0.296 \\
 & 160 & 0.314 & 0.311 & 0.319 & 0.330 & 0.316 & 0.322 & 0.311 & 0.313 \\
 & 170 & 0.322 & 0.319 & 0.330 & 0.341 & 0.327 & 0.333 & 0.326 & 0.328 \\
 & 180 & 0.327 & 0.327 & 0.336 & 0.348 & 0.332 & 0.338 & 0.328 & 0.331 \\
 & 190 & 0.338 & 0.334 & 0.345 & 0.354 & 0.340 & 0.345 & 0.335 & 0.337 \\
 & 200 & 0.346 & 0.344 & 0.358 & 0.367 & 0.354 & 0.362 & 0.353 & 0.355 \\
 & 250 & 0.386 & 0.388 & 0.397 & 0.411 & 0.389 & 0.403 & 0.387 & 0.394 \\
 & 300 & 0.442 & 0.438 & 0.445 & 0.455 & 0.433 & 0.448 & 0.429 & 0.436 \\
 & 350 & 0.490 & 0.485 & 0.494 & 0.502 & 0.486 & 0.494 & 0.480 & 0.480 \\
 & 400 & 0.514 & 0.510 & 0.521 & 0.530 & 0.512 & 0.526 & 0.506 & 0.511 \\
 & 450 & 0.555 & 0.551 & 0.558 & 0.567 & 0.541 & 0.558 & 0.537 & 0.543 \\
 & 500 & 0.586 & 0.587 & 0.588 & 0.597 & 0.572 & 0.589 & 0.566 & 0.576 \\
\midrule
\multirow[t]{26}{*}{17} & 10 & 0.086 & 0.076 & 0.085 & 0.073 & 0.087 & 0.078 & 0.085 & 0.076 \\
 & 20 & 0.111 & 0.108 & 0.111 & 0.105 & 0.111 & 0.105 & 0.111 & 0.106 \\
 & 30 & 0.131 & 0.128 & 0.134 & 0.129 & 0.133 & 0.126 & 0.132 & 0.128 \\
 & 40 & 0.147 & 0.147 & 0.149 & 0.150 & 0.148 & 0.144 & 0.146 & 0.145 \\
 & 50 & 0.159 & 0.157 & 0.164 & 0.165 & 0.162 & 0.158 & 0.164 & 0.162 \\
 & 60 & 0.174 & 0.173 & 0.174 & 0.179 & 0.174 & 0.172 & 0.175 & 0.174 \\
 & 70 & 0.190 & 0.187 & 0.191 & 0.194 & 0.186 & 0.185 & 0.188 & 0.187 \\
 & 80 & 0.203 & 0.201 & 0.205 & 0.211 & 0.204 & 0.203 & 0.204 & 0.203 \\
 & 90 & 0.211 & 0.210 & 0.219 & 0.227 & 0.217 & 0.218 & 0.218 & 0.216 \\
 & 100 & 0.223 & 0.220 & 0.228 & 0.233 & 0.228 & 0.227 & 0.226 & 0.225 \\
 & 110 & 0.230 & 0.229 & 0.236 & 0.242 & 0.231 & 0.234 & 0.232 & 0.231 \\
 & 120 & 0.242 & 0.240 & 0.248 & 0.255 & 0.245 & 0.246 & 0.241 & 0.242 \\
 & 130 & 0.251 & 0.248 & 0.258 & 0.268 & 0.257 & 0.258 & 0.252 & 0.253 \\
 & 140 & 0.262 & 0.259 & 0.266 & 0.279 & 0.265 & 0.269 & 0.263 & 0.264 \\
 & 150 & 0.273 & 0.271 & 0.281 & 0.292 & 0.278 & 0.281 & 0.275 & 0.277 \\
 & 160 & 0.281 & 0.280 & 0.292 & 0.300 & 0.289 & 0.291 & 0.285 & 0.282 \\
 & 170 & 0.289 & 0.285 & 0.298 & 0.308 & 0.287 & 0.297 & 0.285 & 0.289 \\
 & 180 & 0.296 & 0.298 & 0.302 & 0.315 & 0.301 & 0.309 & 0.299 & 0.302 \\
 & 190 & 0.303 & 0.304 & 0.310 & 0.321 & 0.305 & 0.313 & 0.303 & 0.304 \\
 & 200 & 0.307 & 0.308 & 0.318 & 0.330 & 0.311 & 0.320 & 0.309 & 0.311 \\
 & 250 & 0.363 & 0.363 & 0.367 & 0.381 & 0.361 & 0.372 & 0.359 & 0.363 \\
 & 300 & 0.404 & 0.401 & 0.407 & 0.422 & 0.401 & 0.413 & 0.397 & 0.401 \\
 & 350 & 0.439 & 0.439 & 0.447 & 0.459 & 0.439 & 0.453 & 0.434 & 0.438 \\
 & 400 & 0.479 & 0.475 & 0.482 & 0.492 & 0.469 & 0.482 & 0.465 & 0.469 \\
 & 450 & 0.508 & 0.510 & 0.512 & 0.523 & 0.500 & 0.516 & 0.496 & 0.501 \\
 & 500 & 0.538 & 0.540 & 0.540 & 0.556 & 0.529 & 0.546 & 0.523 & 0.534 \\
\midrule
\multirow[t]{26}{*}{18} & 10 & 0.082 & 0.074 & 0.082 & 0.072 & 0.086 & 0.077 & 0.087 & 0.074 \\
 & 20 & 0.103 & 0.097 & 0.103 & 0.095 & 0.105 & 0.098 & 0.104 & 0.097 \\
 & 30 & 0.123 & 0.123 & 0.124 & 0.121 & 0.122 & 0.119 & 0.122 & 0.119 \\
 & 40 & 0.136 & 0.135 & 0.138 & 0.138 & 0.134 & 0.131 & 0.138 & 0.137 \\
 & 50 & 0.148 & 0.148 & 0.149 & 0.152 & 0.149 & 0.145 & 0.150 & 0.148 \\
 & 60 & 0.165 & 0.164 & 0.168 & 0.171 & 0.165 & 0.163 & 0.165 & 0.166 \\
 & 70 & 0.175 & 0.176 & 0.180 & 0.188 & 0.179 & 0.180 & 0.178 & 0.180 \\
 & 80 & 0.193 & 0.190 & 0.192 & 0.197 & 0.190 & 0.188 & 0.190 & 0.188 \\
 & 90 & 0.194 & 0.195 & 0.195 & 0.206 & 0.194 & 0.195 & 0.192 & 0.194 \\
 & 100 & 0.204 & 0.204 & 0.208 & 0.217 & 0.204 & 0.205 & 0.203 & 0.204 \\
 & 110 & 0.216 & 0.214 & 0.217 & 0.224 & 0.215 & 0.217 & 0.214 & 0.215 \\
 & 120 & 0.223 & 0.223 & 0.228 & 0.238 & 0.223 & 0.225 & 0.224 & 0.224 \\
 & 130 & 0.231 & 0.231 & 0.236 & 0.246 & 0.233 & 0.236 & 0.231 & 0.232 \\
 & 140 & 0.242 & 0.243 & 0.248 & 0.255 & 0.243 & 0.245 & 0.242 & 0.242 \\
 & 150 & 0.255 & 0.254 & 0.259 & 0.269 & 0.258 & 0.261 & 0.255 & 0.254 \\
 & 160 & 0.262 & 0.263 & 0.266 & 0.277 & 0.264 & 0.269 & 0.260 & 0.263 \\
 & 170 & 0.266 & 0.265 & 0.273 & 0.282 & 0.273 & 0.275 & 0.271 & 0.269 \\
 & 180 & 0.281 & 0.278 & 0.281 & 0.294 & 0.275 & 0.283 & 0.275 & 0.277 \\
 & 190 & 0.287 & 0.286 & 0.294 & 0.305 & 0.289 & 0.295 & 0.287 & 0.288 \\
 & 200 & 0.294 & 0.294 & 0.299 & 0.310 & 0.296 & 0.301 & 0.294 & 0.293 \\
 & 250 & 0.334 & 0.332 & 0.341 & 0.353 & 0.335 & 0.343 & 0.331 & 0.333 \\
 & 300 & 0.367 & 0.365 & 0.371 & 0.385 & 0.364 & 0.378 & 0.360 & 0.365 \\
 & 350 & 0.408 & 0.405 & 0.407 & 0.421 & 0.398 & 0.411 & 0.396 & 0.399 \\
 & 400 & 0.437 & 0.436 & 0.436 & 0.450 & 0.429 & 0.444 & 0.424 & 0.428 \\
 & 450 & 0.468 & 0.465 & 0.469 & 0.480 & 0.461 & 0.472 & 0.458 & 0.458 \\
 & 500 & 0.499 & 0.498 & 0.504 & 0.512 & 0.493 & 0.504 & 0.490 & 0.491 \\
\midrule
\multirow[t]{26}{*}{19} & 10 & 0.085 & 0.073 & 0.083 & 0.071 & 0.084 & 0.075 & 0.083 & 0.071 \\
 & 20 & 0.102 & 0.098 & 0.102 & 0.095 & 0.101 & 0.095 & 0.101 & 0.096 \\
 & 30 & 0.118 & 0.116 & 0.118 & 0.113 & 0.118 & 0.111 & 0.117 & 0.112 \\
 & 40 & 0.135 & 0.133 & 0.132 & 0.132 & 0.130 & 0.126 & 0.131 & 0.127 \\
 & 50 & 0.140 & 0.140 & 0.139 & 0.141 & 0.138 & 0.135 & 0.140 & 0.139 \\
 & 60 & 0.156 & 0.154 & 0.158 & 0.157 & 0.156 & 0.150 & 0.154 & 0.152 \\
 & 70 & 0.161 & 0.162 & 0.164 & 0.169 & 0.163 & 0.161 & 0.165 & 0.163 \\
 & 80 & 0.176 & 0.178 & 0.176 & 0.183 & 0.177 & 0.176 & 0.178 & 0.178 \\
 & 90 & 0.188 & 0.186 & 0.189 & 0.193 & 0.187 & 0.186 & 0.185 & 0.184 \\
 & 100 & 0.196 & 0.195 & 0.199 & 0.205 & 0.196 & 0.196 & 0.198 & 0.194 \\
 & 110 & 0.208 & 0.207 & 0.214 & 0.218 & 0.208 & 0.206 & 0.207 & 0.204 \\
 & 120 & 0.211 & 0.208 & 0.212 & 0.221 & 0.210 & 0.211 & 0.211 & 0.211 \\
 & 130 & 0.221 & 0.219 & 0.223 & 0.232 & 0.221 & 0.221 & 0.218 & 0.217 \\
 & 140 & 0.224 & 0.224 & 0.235 & 0.241 & 0.231 & 0.230 & 0.228 & 0.225 \\
 & 150 & 0.237 & 0.236 & 0.241 & 0.248 & 0.240 & 0.240 & 0.239 & 0.236 \\
 & 160 & 0.241 & 0.239 & 0.245 & 0.254 & 0.244 & 0.246 & 0.241 & 0.239 \\
 & 170 & 0.246 & 0.249 & 0.255 & 0.266 & 0.253 & 0.255 & 0.249 & 0.249 \\
 & 180 & 0.256 & 0.255 & 0.263 & 0.271 & 0.259 & 0.264 & 0.255 & 0.255 \\
 & 190 & 0.267 & 0.267 & 0.269 & 0.280 & 0.266 & 0.272 & 0.264 & 0.265 \\
 & 200 & 0.267 & 0.267 & 0.273 & 0.286 & 0.270 & 0.276 & 0.266 & 0.268 \\
 & 250 & 0.308 & 0.306 & 0.315 & 0.324 & 0.309 & 0.313 & 0.309 & 0.305 \\
 & 300 & 0.340 & 0.338 & 0.344 & 0.358 & 0.335 & 0.346 & 0.333 & 0.337 \\
 & 350 & 0.369 & 0.371 & 0.373 & 0.387 & 0.366 & 0.377 & 0.363 & 0.367 \\
 & 400 & 0.405 & 0.403 & 0.407 & 0.419 & 0.400 & 0.411 & 0.398 & 0.397 \\
 & 450 & 0.437 & 0.435 & 0.440 & 0.450 & 0.428 & 0.438 & 0.425 & 0.424 \\
 & 500 & 0.463 & 0.463 & 0.464 & 0.474 & 0.446 & 0.460 & 0.445 & 0.446 \\
\midrule
\multirow[t]{26}{*}{20} & 10 & 0.079 & 0.068 & 0.078 & 0.065 & 0.077 & 0.069 & 0.079 & 0.066 \\
 & 20 & 0.098 & 0.094 & 0.098 & 0.091 & 0.102 & 0.094 & 0.100 & 0.092 \\
 & 30 & 0.114 & 0.113 & 0.116 & 0.112 & 0.116 & 0.110 & 0.117 & 0.111 \\
 & 40 & 0.128 & 0.126 & 0.126 & 0.125 & 0.128 & 0.122 & 0.130 & 0.123 \\
 & 50 & 0.142 & 0.139 & 0.140 & 0.140 & 0.139 & 0.134 & 0.141 & 0.137 \\
 & 60 & 0.146 & 0.145 & 0.147 & 0.148 & 0.146 & 0.143 & 0.148 & 0.146 \\
 & 70 & 0.154 & 0.153 & 0.156 & 0.157 & 0.155 & 0.152 & 0.154 & 0.152 \\
 & 80 & 0.161 & 0.162 & 0.164 & 0.168 & 0.161 & 0.160 & 0.163 & 0.161 \\
 & 90 & 0.171 & 0.170 & 0.173 & 0.176 & 0.171 & 0.169 & 0.171 & 0.169 \\
 & 100 & 0.179 & 0.178 & 0.181 & 0.185 & 0.180 & 0.176 & 0.183 & 0.178 \\
 & 110 & 0.192 & 0.192 & 0.195 & 0.201 & 0.194 & 0.194 & 0.193 & 0.192 \\
 & 120 & 0.196 & 0.196 & 0.197 & 0.206 & 0.197 & 0.198 & 0.197 & 0.196 \\
 & 130 & 0.209 & 0.207 & 0.212 & 0.218 & 0.210 & 0.207 & 0.208 & 0.205 \\
 & 140 & 0.218 & 0.214 & 0.218 & 0.224 & 0.214 & 0.215 & 0.213 & 0.212 \\
 & 150 & 0.218 & 0.220 & 0.221 & 0.232 & 0.218 & 0.221 & 0.218 & 0.219 \\
 & 160 & 0.229 & 0.227 & 0.236 & 0.242 & 0.232 & 0.231 & 0.229 & 0.226 \\
 & 170 & 0.232 & 0.232 & 0.234 & 0.244 & 0.234 & 0.236 & 0.229 & 0.230 \\
 & 180 & 0.236 & 0.236 & 0.242 & 0.249 & 0.236 & 0.238 & 0.235 & 0.234 \\
 & 190 & 0.252 & 0.249 & 0.256 & 0.263 & 0.253 & 0.253 & 0.250 & 0.246 \\
 & 200 & 0.257 & 0.256 & 0.254 & 0.264 & 0.252 & 0.257 & 0.251 & 0.251 \\
 & 250 & 0.284 & 0.285 & 0.292 & 0.303 & 0.288 & 0.292 & 0.283 & 0.283 \\
 & 300 & 0.317 & 0.315 & 0.320 & 0.333 & 0.314 & 0.322 & 0.310 & 0.310 \\
 & 350 & 0.349 & 0.347 & 0.354 & 0.362 & 0.343 & 0.350 & 0.340 & 0.337 \\
 & 400 & 0.384 & 0.382 & 0.386 & 0.397 & 0.373 & 0.383 & 0.368 & 0.370 \\
 & 450 & 0.401 & 0.403 & 0.403 & 0.414 & 0.397 & 0.406 & 0.394 & 0.392 \\
 & 500 & 0.427 & 0.430 & 0.421 & 0.439 & 0.409 & 0.428 & 0.406 & 0.414 \\
\midrule
\end{longtable}

\section{Analytical Derivation of Kullback--Leibler Divergence}\label{sec:derivation_DKL}

We derive an exact expression for the Kullback--Leibler (KL) divergence between the finite-$N$ velocity distribution $p_N(x)$ and its limit, the Gaussian distribution $p_\infty(x)$.

The divergence is defined as:
\begin{equation}
    D_{\mathrm{KL}}(p_N \| p_\infty) = \int_{-\sqrt{N}}^{\sqrt{N}} p_N(x) \ln p_N(x) \, dx - \int_{-\sqrt{N}}^{\sqrt{N}} p_N(x) \ln p_\infty(x) \, dx.
\end{equation}
The Gaussian term (second term) evaluates to:
\begin{equation}
    - \mathbb{E}_{p_N} \left[ -\frac{1}{2}\ln(2\pi) - \frac{x^2}{2} \right] = \frac{1}{2}(1 + \ln 2\pi),
\end{equation}
since the variance of $p_N$ is constrained to be unity.

The entropy term (first term) involves the expectation of $\ln(1 - x^2/N)$. Using the substitution $t = x/\sqrt{N}$ and the properties of the Beta function derivative, specifically $\partial_y B(x,y) = B(x,y)(\psi(y) - \psi(x+y))$, we find:
\begin{equation}
    \int_{-\sqrt{N}}^{\sqrt{N}} p_N(x) \ln \left(1 - \frac{x^2}{N}\right) dx = \psi\left(\frac{N-1}{2}\right) - \psi\left(\frac{N}{2}\right),
\end{equation}
where $\psi(z)$ is the digamma function.
Combining these results, we obtain the closed-form solution:
\begin{equation}
    D_{\mathrm{KL}}(p_N \| p_\infty) = \ln C_N + \frac{1}{2}(1 + \ln 2\pi) + \frac{N-3}{2} \left[ \psi\left(\frac{N-1}{2}\right) - \psi\left(\frac{N}{2}\right) \right],
\end{equation}
with $C_N = \frac{\Gamma(N/2)}{\sqrt{N\pi}\,\Gamma((N-1)/2)}$. This formula allows for precise calculation of the information-theoretic distinguishability limit.

\section{Derivation of the Continuous Probability Functional from the Discrete}
\label{sec:sanov_functional}
We derive the continuous analog of the discrete probability $\mathbb{P}_Q(P)$ for observing an empirical distribution $P$ given a true distribution $Q$.

\subsection{Discretization Setup}
Consider a continuous domain $\Omega$ partitioned into $k$ infinitesimal bins of width $\Delta x$. Let $q(x)$ be the true probability density function and $p(x)$ be the observed empirical density. The discrete probabilities for the $i$-th bin located at $x_i$ are given by:
\begin{equation}
    q_i \approx q(x_i) \Delta x, \quad p_i \approx p(x_i) \Delta x.
\end{equation}
The number of particles in the $i$-th bin is $n_i = n p_i = n p(x_i) \Delta x$.

\subsection{Log-Probability Expansion}
Starting from the discrete multinomial formula:
\begin{equation}
    \mathbb{P}_Q(P) = \frac{n!}{\prod_{i=1}^k n_i!} \prod_{i=1}^k q_i^{n_i}.
\end{equation}
Taking the natural logarithm:
\begin{equation}
    \ln \mathbb{P}_Q(P) = \ln n! - \sum_{i=1}^k \ln (n_i!) + \sum_{i=1}^k n_i \ln q_i.
\end{equation}
Applying Stirling's approximation ($\ln n! \approx n \ln n - n$) for large $n$:
\begin{align}
    \ln \mathbb{P}_Q(P) &\approx (n \ln n - n) - \sum_{i=1}^k (n_i \ln n_i - n_i) + \sum_{i=1}^k n_i \ln q_i \nonumber \\
    &= n \ln n - \sum_{i=1}^k n p_i \ln (n p_i) + \sum_{i=1}^k n p_i \ln q_i \nonumber \\
    &= - n \sum_{i=1}^k p_i \ln \frac{p_i}{q_i}.
\end{align}

\subsection{The Continuum Limit}
Now, we substitute the continuous forms $p_i = p(x_i)\Delta x$ and $q_i = q(x_i)\Delta x$ into the logarithmic term inside the summation:
\begin{align}
    \ln \frac{p_i}{q_i} &= \ln \left( \frac{p(x_i) \Delta x}{q(x_i) \Delta x} \right) = \ln \left( \frac{p(x_i)}{q(x_i)} \right).
\end{align}
Notice that the discretization width $\Delta x$ cancels out inside the logarithm. The sum then transforms into an integral as $\Delta x \to 0$:
\begin{equation}
    \sum_{i=1}^k p_i \ln \frac{p_i}{q_i} = \sum_{i=1}^k (p(x_i) \Delta x) \ln \frac{p(x_i)}{q(x_i)} \xrightarrow{\Delta x \to 0} \int p(x) \ln \frac{p(x)}{q(x)} \, dx.
\end{equation}

\subsection{Result}
Substituting this back into the expression for $\ln \mathbb{P}_Q(P)$, we obtain the continuous probability functional:
\begin{equation}
    \ln \mathbb{P}_q(p) \approx -n \int p(x) \ln \frac{p(x)}{q(x)} \, dx = -n D_{\mathrm{KL}}(p \| q).
\end{equation}
Exponentiating gives the final continuous form:
\begin{equation}
    \mathbb{P}_q(p) \propto \exp \left( -n D_{\mathrm{KL}}(p \| q) \right).
\end{equation}
This derivation confirms that the Kullback--Leibler divergence is the natural rate function for the probability density in the continuous domain, emerging directly from the limit of the discrete multinomial distribution.

\end{appendices}

\bibliography{references}

\end{document}